\documentclass[10pt,letterpaper]{article}
\usepackage[margin=0.9in]{geometry}
\usepackage{amsmath,amssymb}
\usepackage{mathtools}

\usepackage{changepage}

\usepackage[utf8x]{inputenc}

\usepackage{textcomp,marvosym}

\usepackage{cite}

\usepackage[right]{lineno}

\usepackage{microtype}
\DisableLigatures[f]{encoding = *, family = * }

\usepackage[table,dvipsnames]{xcolor}

\usepackage{array}

\usepackage{soul}
\usepackage{subfigure}

\usepackage{tablefootnote}

\newcolumntype{+}{!{\vrule width 2pt}}

\newlength\savedwidth

\usepackage[aboveskip=1pt,labelfont=bf,labelsep=period,justification=raggedright,singlelinecheck=off]{caption}

\makeatletter
\renewcommand{\@biblabel}[1]{\quad#1.}
\makeatother

\date{}

\usepackage{lastpage,fancyhdr,graphicx,dsfont}
\usepackage{comment}
\usepackage{epstopdf}

\def\0{{\bf 0}}
\def\1{{\bf 1}}
\def\2{{\bf 2}}
\def\3{{\bf 3}}
\def\4{{\bf 4}}

\def\dd{\textrm{d}}

\def\cd{\color{red}}
\usepackage{changes}

\usepackage[dvipsnames]{xcolor}

\global\long\def\P#1{\mathbb{P}\left[#1\right]}%
\global\long\def\E#1{\mathbb{E}\left[#1\right]}%

\global\long\def\ind#1{\mathds{1}\left(#1\right)}%

\usepackage{xcolor}

\usepackage{subfig}

\begin{document}

\vspace*{0.2in}

\begin{flushleft}
{\Large \textbf\newline{{How heterogeneous thymic output and 
homeostatic proliferation shape naive T cell receptor clone abundance distributions}}}
%
%
%
\newline
\\
Renaud Dessalles\textsuperscript{1}, Yunbei Pan\textsuperscript{1,2},
Mingtao Xia\textsuperscript{3},
Davide Maestrini\textsuperscript{1}  
Maria R. D'Orsogna\textsuperscript{1,2},
Tom Chou\textsuperscript{1,3},
\\
\bigskip
\textbf{1} Dept. of Computational Medicine, UCLA, Los Angeles, CA 90095-1766
\\
\textbf{2} Dept. of Mathematics, CalState-Northridge, Los Angeles, CA 91330
\\
\textbf{3} Dept. of Mathematics, UCLA, Los Angeles, CA 90095-1555  
$\ast$ E-mail: tomchou@ucla.edu
\\
\bigskip


\end{flushleft}

\section*{Abstract}

The set of T cells that express the same T cell receptor (TCR)
sequence represents a T cell clone.  The number of different naive T
cell clones in an organism reflects the number of different T cell
receptors (TCRs) arising from recombination of the V(D)J gene segments
during T cell development in the thymus.  TCR diversity and more
specifically, the clone abundance distribution is an important factor
in immune function.  Specific recombination patterns occur more
frequently than others while subsequent interactions between TCRs and
self-antigens are known to trigger proliferation and sustain naive T
cell survival. These processes are TCR-dependent, leading to
clone-dependent thymic export and naive T cell proliferation rates.
Using a mean-field approximation to the solution of a regulated
birth-death-immigration model and a modification arising from
sampling, we systematically quantify how TCR-dependent heterogeneities
in immigration and proliferation rates affect the shape of clone
abundance distributions (the number of different clones that are
represented by a specific number of cells, or ``clone counts'').  By
comparing predicted clone counts derived from our heterogeneous
birth-death-immigration model with experimentally sampled clone
abundances, we show that although heterogeneity in immigration rates
causes very little change to predicted clone-counts, significant
heterogeneity in proliferation rates is necessary to generate the
observed abundances with reasonable physiological parameter values.
Our analysis provides constraints among physiological parameters that
are necessary to yield predictions that qualitatively match the
data. Assumptions of the model and potentially other important
mechanistic factors are discussed.

\section*{Author Summary}
It is not known how naive T cell receptor (TCR)-dependent
  thymic output, proliferation, and death rates control the overall
  diversity of T cells in the body.  Using a reasonable range of
  physiological parameters within a birth-death-immigration (BDI)
  model, we quantitatively show how heterogeneity influences the
  expected sampled clone abundances. We find that heterogeneity in
  clone-specific immigration rate only weakly affects sampled clone
  counts while modest heterogeneity in proliferation rates can
  dramatically affect clone counts. Fitting our model to recently
  published data, we impose quantitative constraints on biologically
  meaningful rate parameters in the model. These findings are
  consistent with the known proliferation-driven maintenance of T cell
  population in humans.

\section*{Introduction}

Naive T cells play a fundamental role in the immune system's response
to pathogens, tumors, and other infectious agents. These cells are
produced in the thymus, circulate through the blood, and migrate to
the lymph nodes where they may be presented with different antigen
proteins from various pathogens. Naive T cells mature in the thymus
where the so-called V, D, and J segments of genes that code T cell
receptors undergo rearrangement. Most T cell receptors (TCRs) are
comprised of an alpha chain and a beta chain that are formed after VJ
segment and VDJ segment recombination, respectively. The number of
possible TCR gene sequences is extremely large, but while
recombination is a nearly random process, not all TCRs are formed with
the same probability. Before export to the periphery, T cells undergo
a selection process, during which T cells with TCRs that react to self
proteins are eliminated.

The unique receptors expressed on the cell surface of circulating TCRs
enable them to recognize specific antigens; well-known examples
include the naive forms of helper T cells (CD4+) and cytotoxic T cells
(CD8+).  The set of naive T cells that express the same TCR are said
to belong to the same T cell clone.  Upon encountering the antigens
that activate their TCRs, naive T cells turn into effector cells that
assist in eliminating infected cells. Effector cells die after
pathogen clearance, but some develop into memory T cells.  Because of
the large number of unknown pathogens, TCR clonal diversity is a key
factor for mounting an effective immune response.  Recent studies also
reveal that human TCR clonal diversity is implicated in healthy aging,
neonatal immunity, vaccination response and T cell reconstitution
following haematopoietic stem cell transplantation
\cite{Qi2014,VandenBroek2018}.  Despite the central role of the naive
T cell pool in host defense, and broadly speaking in health and
disease, TCR diversity is difficult to quantify.  For example, the
human body hosts a large repertoire of T cell clones, however the
actual distribution of clone sizes is not precisely known
\cite{Laydon2015}.  Only recently have experimental and theoretical
efforts been devoted to understanding the mechanistic origins of TCR
diversity
\cite{Desponds2016,Desponds2017,Lythe2016,deGreef2018,Koch2018,denBraber2012}.
The goal of this work is to formulate a realistic mathematical model
that includes heterogeneous naive T cell generation and reproduction
rates and that we will use to describe recent experimental results.

A well-established way to describe the T cell repertoire is by
determining the clone abundance distribution or ``clone count''
$\hat{c}_{k}$ (for $k\geq1$) that measures the number of distinct
clones represented by exactly $k$ T cells: $\hat{c}_{k} \equiv
\sum_{i=1}^{\infty}\mathds{1}(n_{i},k)$, where $n_{i}$ is the discrete
number of T cells carrying TCR $i$ and the indicator function
$\mathds{1}(n,k) = 1$ if $n=k$ and $0$ otherwise. This distribution
captures the entire pattern of the clonal populations.  Several
summary indices for T cell diversity such as Shannon's entropy,
Simpson's index, or the whole population richness can be deduced from
the distribution $\hat{c}_{k}$ \cite{Dessalles2018}. Note that
$\hat{c}_k$ counts only the number of clones of a specific population
$k$ and does not carry any TCR sequence or identity information.

\begin{figure}[h!]
\centering{}
\includegraphics[scale=0.75]{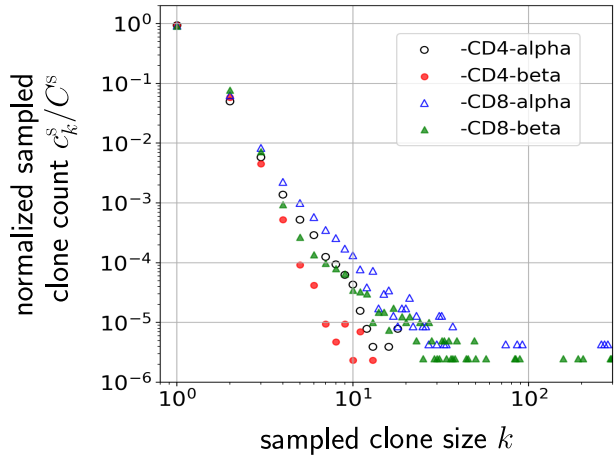}
\vspace{2mm}
\caption{Normalized naive T cell clone count data sampled from one
  patient in Oakes \textit{et al}. \cite{Oakes2017} plotted on a
  log-log scale. The clones counts plotted are the average of three
  samples among CD4 and CD8 cell subgroups. Clones are defined by
  different nucleotide sequences associated with different alpha or
  beta chains of the TCR.}
     \label{DATA}
\end{figure}

Complete clone counts $\hat{c}_k$ and the total number of circulating
naive T cells are difficult to measure in humans. Nonetheless,
high-throughput DNA sequencing on samples of peripheral blood
containing T cells
\cite{Mora2010,Oakes2017,Aguilera-Sandoval2016,Gerritsen2016} have
provided some insight into TCR diversity.  A commonly observed feature
is that clone counts $\hat{c}_k$ often exhibit a power-law
distribution in the clone abundance $k$, albeit for approximately one
or at most two decades (see Fig.~\ref{DATA}).

Several models have been developed to explain certain features of
observed clone counts
\cite{Laydon2015,Burgos1996,Weinstein2009,Naumov2003,Desponds2016},
including the apparent power-law behavior.  One proposal is that T
cells in different clones have TCRs that have different affinities for
self-ligands that are necessary for peripheral proliferation
\cite{Desponds2016,Desponds2017,Lythe2016}, leading to clone specific
replication rates. An alternative hypothesis \cite{deGreef2018} is
that specific TCR sequences are more likely to arise in the V(D)J
recombination process in the thymus \cite{Marcou2018} leading to a
higher probability that these TCRs are produced.  De Greef \emph{et
  al.}  \cite{deGreef2018} estimated the probability of production of
a given TCR sequence by using the Inference and Generation of
Repertoires (IGoR) simulation tool that quantitatively characterizes
the statistics of receptor generation from both cDNA and gDNA data
\cite{Marcou2018}. However, the data indicate that power laws, which
represent a simple summary of sampled clone counts, arise only across
one or two decades of clone sizes, as shown in
Fig.~\ref{DATA}. Moreover, the above studies have not systematically
incorporated and compared heterogeneity in both immigration and
replication rates, sampling, or fitting to measured TCR clone
abundance distributions.

In this paper, we analyze the effects of heterogeneity and sampling
within a mean-field model based on a stochastic multiclone
birth-death-immigration (BDI) process. The model allows for
TCR-sequence dependent replication and immigration rates.  Our model
is derived from an underlying continuous-time Markovian
birth-death-immigration (BDI) process \cite{Dessalles2018} described
by: (i)~immigration representing the arrival of new clones from the
thymus; (ii)~birth during homeostatic proliferation of naive T cells
that yield newborn naive T cells with the same TCR as their parent;
and (iii)~death representing cell apoptosis.  We also include a
regulation, or ``carrying capacity,'' mechanism through a total
population-dependent death rate which may represent the global
competition for cytokines, such as Interleukine-7
\cite{Tan2001,Schluns2000,Ciupe2009,LYTHE_IL7,SOUSA2016}, needed for
naive T cell survival and homeostasis \cite{SURH2008,FARBER2014}.
This interaction will ensure a homeostatic finite naive T cell
population and will be assumed to be clone-independent since these
cytokine signals are TCR-independent \cite{Ciupe2009}.

We derive analytical results of our heterogeneous BDI model that are
applicable on the scale of the entire organism. These results are then
extended to predict the expected clone counts in a small subsample of
the population. Finally, this result is averaged over distributions of
TCR immigration (thymic output) and homeostatic proliferation
rates. Heterogeneity in TCR production rates are extracted from new
computational tools: Inference and Generation of Repertoires (IGoR)
\cite{Marcou2018} and Optimized Likelihood estimate of immunoGlobulin
Amino-acid sequences (OLGA) \cite{OLGA}. Since there are no equivalent
tools that measures proliferation rates, we will assume simple
functional forms for the distribution of homeostatic proliferation
rates.

Our calculations provide insight into how parameters describing the
shape of the distribution of immigration and proliferation rates
affect the shape of the expected clone counts. To compare with
experimental measurements, we also quantify the random sampling
process that describes actual measurements derived from blood draws.
Our model allows us to estimate the values of relevant physiological
parameters and reveals that the shape of the sampled expected clone
counts are much less sensitive to TCR-dependent thymic output than to
TCR-dependent proliferation.  However, we show how the width of a
simple uniform proliferation rate distribution can inform the observed
$\hat{c}_{k}$.


\section*{Materials and Methods}

To understand the observed clone counts, we focus on the clone
abundance distribution $\hat{c}_k$ associated only with naive T cells,
the first type of cells produced by the thymus that have not yet been
activated by any antigen. Antigen-mediated activation initiates a
one-way irreversible cascade of differentiation into effector and
memory T cells that we can subsume into a death rate.  Thus, we limit
our analysis to birth, death, and immigration within the naive T cell
compartment. Here, we first present the mathematical framework of the
BDI process to provide an initial qualitative understanding for clone
counts.

\subsection*{Non-neutral Birth-Death-Immigration model}

\begin{figure}[h!]
\centering{}
\includegraphics[scale=0.7]{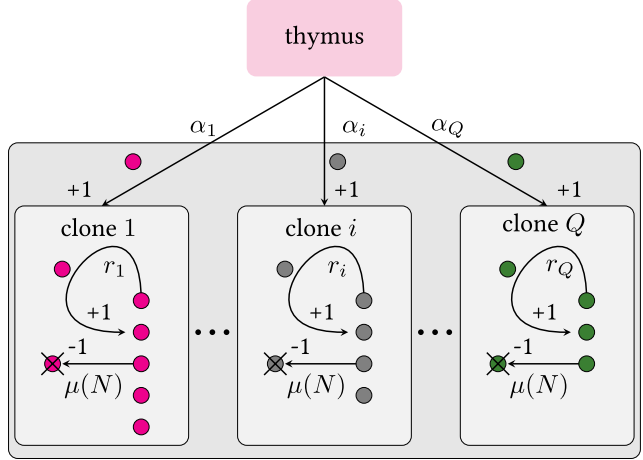}
\vspace{2mm}
\caption{Schematic of a multiclone birth-death-immigration
  process. Clones are defined by distinct TCR sequences $i$.  Each
  clone carries its own thymic output and peripheral proliferation
  rates, $\alpha_{i}$ and $r_{i}$, respectively. We assume all clones
  have the same population-dependent death rate $\mu(N)$. Since $Q \gg
  1$, we impose a continuous distribution over the rates $\alpha$ and
  $r$. Theoretically, there may be $Q \gtrsim 10^{15}$
  \cite{Lythe2016} or more \cite{ROBINS2009,MAYER_PNAS_2019} possible
  viable V(D)J recombinations. The actual, effective number of
  different selected TCRs sequences is expected to be much less since
  extremely low probability sequences may never be formed during the
  organism's lifetime. A strict lower bound on $Q$ is the actual
  number of clones predicted in an entire organism ($\sim 10^{7}$ for
  human)
  \cite{ARSTILA1999,WARREN2011,Zarnitsyna2013,Qi2014,Lythe2016}.
     \label{Fig1}}
\end{figure}

The multiclone BDI process is depicted in Fig.~\ref{Fig1}.
We define $Q$ to be the theoretical number of possible functional
naive T cell receptor clones that can be generated by V(D)J
recombination in the thymus which is estimated to be $Q\sim
10^{13}-10^{18}$ \cite{Lythe2016,OLGA}.  As we will later show,
results of our model will not be depend the explicit value of $Q$ as
long as $Q\gg 1$.  Due to naive T cell death or removal from the
sampling-accessible pool, not all possible clone types will be
presented in the organism, so we denote the number of clones actually
present in the body (or ``richness'') by $C \ll Q$, where estimates of
$C$ range from $\sim 10^{6}-10^{8}$ in mice and humans
\cite{ARSTILA1999,WARREN2011,Qi2014,Lythe2016,Jenkins2010}.

Although naive T cells are difficult to distinguish from the entire T
cell population, the total number of naive T cells (across all clones
present) in human has been estimated to be about $N \sim
10^{11}$. However, circulating naive T cells number perhaps $10^9$ but
can exchange, at different time scales, with those that reside in
peripheral tissue, which may carry their own proliferation and death
rates. The \textit{effective} pool that is ultimately sampled is thus
difficult to estimate, but measurements show the theoretical number of
different clones $\gg$ the total number of naive T cells $\gg$ the
total number of different T cells clones actually in the body ($Q \gg
N \gg C$). Regardless of the precise values of the discrete quantities
$Q, C, N$, they are related to the discrete clone counts $\hat{c}_k$
through

\begin{equation}
C=\sum_{k\geq1}\hat{c}_k = Q-\hat{c}_{0}\qquad\text{and}\qquad N=\sum_{k\geq1}k \hat{c}_k,
\label{RELATIONSHIPS}
\end{equation}
where $\hat{c}_0$ is the number of possible clones that are not
expressed in the organism.

Each distinct clone $i$ (with $1\leq i \leq Q$) is characterized by an
immigration rate $\alpha_i$ and a per cell replication rate $r_{i}$.
The immigration rate $\alpha_i$ is clone-specific because it depends
on the preferential V(D)J recombination process; the replication rate
$r_{i}$ is also clone-specific due to the different interactions with
self-peptides that trigger proliferation. Since both the numbers of
theoretically possible ($Q\gg 1$) and observed ($C \gg 1$) clones are
extremely large, we can define a continuous density $\pi(\alpha,r)$
from which immigration and proliferation rates $\alpha$ and $r$ are
drawn.  This means that the probability that a given clone has an
immigration rate between $\alpha$ and $\alpha + \dd \alpha$ and
replication rate between $r$ and $r+\dd r$ is $\pi(\alpha,r)
\dd\alpha\dd r$.  Since $Q$ is finite and countable, there are
theoretical maximum values $A$ and $R$ for the immigration and
proliferation rates, respectively, such that $\pi(\alpha,r) = 0$ for
$\alpha \geq A$ and $r \geq R$.  In the BDI process, the upper bound
$R$ on the proliferation rate prevents unbounded numbers of naive T
cells and is necessary for a self-consistent solution.  Conversely,
results are rather insensitive to the upper bound $A$ on the
immigration rate provided $\pi(\alpha,r)$ decays sufficiently fast
such that the mean value of $\alpha$ is finite for all $r \leq
R$. Thus, for simplicity, we henceforth take the $A \to \infty$ limit.
The heterogeneity in the immigration and replication rates allows us
to go beyond typical ``neutral'' BDI models, where both rates are
fixed to a specific value for all clones.

Finally, we assume the per cell death rate $\mu(N)$ is
clone-independent but a function of the total population $N$.  This
dependence represents the competition among all naive T cells for a
common resource (such as cytokines), which effectively imposes a
carrying capacity on the population
\cite{LYTHE_IL7,MAYER_PNAS_2019,Lewkiewicz2018}. The specific form of
the regulation will not qualitatively affect our findings since we
will ultimately be interested in only its value $\mu(N^{*}) \equiv
\mu^{*}$ at the steady state population $N^{*}$.

\subsection*{Mean-Field Approximation of the BDI Process}

The exact steady-state probabilities of configurations of the discrete
abundances $\hat{c}_k$ for a fully stochastic neutral BDI model with
regulated death rate $\mu(N)$ were recently derived
\cite{Dessalles2018}. To incorporate TCR-dependent immigration and
replication rates in a non-neutral model, we must consider distinct
values of $\alpha_{i}$ and $r_{i}$ for each clone $i$.  In this case,
an analytic solution for the probability distribution over
$\hat{c}_k$, even at steady state, cannot be expressed in an explicit
form.  However, since the effective number of naive T cells ($N\sim
10^{9}-10^{11}$ \cite{Jenkins2010}) is large, we can exploit a
mean-field approximation to the non-neutral BDI model and derive
expressions for the mean values of the discrete clone counts
$\hat{c}_k$. We will show later that under realistic parameter
regimes, the mean-field approximation is quantitatively
accurate. Breakdown of the mean field approximation has been carefully
analyzed in other studies \cite{Xu2018a}.


\paragraph*{Deterministic approximation for total population and effective death rate:}
To implement the mean-field approximation in the presence of a general
regulated death rate $\mu(N)$, we start by writing the deterministic,
``mass-action'' ODE for the mean density $n_{\alpha,r}(t)$ of cells
having immigration rate near $\alpha$ and proliferation rate near $r$
in a BDI process

\begin{equation}
{\dd n_{\alpha,r}(t)\over \dd t} = \alpha+rn_{\alpha,r}(t)-\mu(N(t))n_{\alpha,r}(t).
\label{eq:nar}
\end{equation}
Since $Q\gg 1$, the total number of clones that carry an immigration
rate between $\alpha$ and $\alpha+\dd\alpha$ and a replication rate
between $r$ and $r+\dd r$ is approximately $Q\pi(\alpha,r)\dd \alpha
\dd r$.  The total mean number $N(t)$ of naive T cells can
then be estimated as a weighted integral over all
$n_{\alpha,r}(t)$

\begin{equation}
N(t)=Q\int_{0}^{\infty}\!\!\!\dd \alpha \!\int_{0}^{R}\!\!\!\dd r\, n_{\alpha,r}(t)
\pi(\alpha,r).
\label{eq:N_n_alpha,r}
\end{equation}
At steady-state, the solution to Eq.~\ref{eq:nar} can be simply expressed as

\begin{equation}
n_{\alpha,r}^{*}=\frac{\alpha}{\mu(N^*)-r}
\label{eq:n_star}
\end{equation}
in which $N^*$ is the steady-state value of $N(t)$.  Thus, upon
averaging Eq.~\ref{eq:n_star} over $\alpha$ and $r$, we find

\begin{equation}
N^*=Q\int_{0}^{R}\!\!\!\dd r \!\int_{0}^{\infty}\!\!\!\dd \alpha \,
\frac{\alpha \pi(\alpha,r)}{\mu(N^*)-r},
\label{eq:fixed-point-equation}
\end{equation}
a self-consistent equation for $N^*$ if the functional form of
$\mu(N^{*})$ is known.  Without the finite upper bound $R$ of the
density $\pi(\alpha,r)$, the integral in
Eq.~\ref{eq:fixed-point-equation} diverges. The self-consistent
determination of $N^{*}$ depends implicitly on the parameters that
define the distribution $\pi(\alpha, r)$.

Note that for factorisable $\pi(\alpha,r)
=\pi_{\alpha}(\alpha)\pi_{r}(r)$, the self-consistent death rate
$\mu^{*}$ depends only on the combination $N^{*}/(\bar{\alpha}Q) =
\int_{0}^{R}\!\dd r\, \pi_{r}(r)/(\mu^{*}-r)$, where $\bar{\alpha}
\equiv \int_{0}^{\infty}\alpha\pi_{\alpha}(\alpha)$ and $\mu^{*}
\equiv \mu(N^{*})$ is the effective death rate \textit{at
  steady-state}. We now nondimensionalize all rates with respect to
the maximum proliferation rate $R=2\bar{r} < \mu^{*}$.  In these
units, $\bar{r}=1/2$ and $0< w \leq 1$. Henceforth, we will use the
notation $\bar{\alpha}/R \to \bar{\alpha}$ and $\mu^{*}/R \to \mu^{*}$
so that the quantities $\bar{\alpha}$ and $\mu^*$ are meant to be
dimensionless, unless otherwise stated. The steady-state
self-consistent condition becomes

\begin{equation}
{N^{*}\over \bar{\alpha}Q} \equiv {\lambda \over \bar{\alpha}}
=\int_{0}^{1}\!\dd r {\pi_{r}(r) \over \mu^{*} - r},
\label{eq:fixed-point-equation_2}
\end{equation}
where $\lambda \equiv N^{*}/Q$ is the dimensionless ratio of the total
naive T cell population to the total effective number of clones
$Q$. We expect $\lambda \ll 1$. For a given $\pi_{r}(r)$, one can then
self-consistently determine $\mu^{*}$ in terms of
$\lambda/\bar{\alpha}$ and the parameters defining $\pi_{r}(r)$.


\paragraph*{Mean-field model of clone counts:}
We now use the results for the mean steady-state population $N^*$ to
find the clone counts averaged over all realizations of the underlying
stochastic process. The mean-field equations for the dynamics of these
mean clone counts in the neutral model were derived in
\cite{GOYAL_BMC,Xu2018a} and are reviewed in Appendix
\ref{app:neutral_model}. In the neutral model, we assume that all
effective clones $Q$ are associated with the same rates $\alpha$ and
$r$ so that the mean field evolution equation for $c_{k}(\alpha, r)$
is \cite{GOYAL_BMC,Xu2018a}


\begin{equation}
{\dd c_k(\alpha,r) \over \dd t} = \alpha\left[c_{k-1}(\alpha,r)
-c_k(\alpha,r)\right] + r\left[(k-1)c_{k-1}(\alpha,r)-kc_k(\alpha,r)\right]
+\mu(N)\left[(k+1)c_{k+1}(\alpha,r)-kc_k(\alpha,r)\right],
\label{CK_ODE}
\end{equation}
along with the constraint $\sum_{k=0}^{\infty}c_{k}(\alpha, r) = Q$.
This equation assumes that both $c_{k}(\alpha, r)$ and $N$ are
uncorrelated, allowing us to write the last term as a product of
functions of the mean population $N=\sum_{k}k c_{k}$ and
$c_{k+1}, c_k$.

At steady state we replace $\mu(N)$ with $\mu(N^*)\equiv \mu^{*}$
self-consistently found by solving
Eq.~\ref{eq:fixed-point-equation_2}.  The solution follows a negative
binomial distribution with parameters $\alpha/r$ and $r/\mu^{*}$
\cite{Dessalles2018,Xu2018a}

\begin{equation}
c_{k}(\alpha, r\vert \mu^{*})=Q\left(1-\frac{r}{\mu^{*}}\right)^{\alpha/r}
\left(\frac{r}{\mu^{*}}\right)^{k}\frac{1}{k!}\prod_{\ell=0}^{k-1}
\left(\frac{\alpha}{r}+\ell\right).
\label{eq:ck_alpha_r}
\end{equation}
The dependence on $\lambda/\bar{\alpha}$ arises from
the determination of $\mu^{*}(\lambda/\bar{\alpha})$
through Eq.~\ref{eq:fixed-point-equation_2}.  Although $c_{k}(\alpha,
r\vert \mu^{*})$ has not yet been averaged over $\alpha, r$, it also
implicitly depends on parameters that define $\pi_{r}(r)$ through the
solution for $\mu^{*}$ from Eq.~\ref{eq:fixed-point-equation_2}.

\subsection*{Sampling}

Unless an animal is sacked and its entire naive T cell population
is sequenced, TCR clone distributions are typically measured
from sequencing TCRs in a small blood sample. In such samples, low
population clones may be missed. In order to compare our predictions
with measured clone abundance distributions, we must revise our
predictions to allow for random cell sampling.

We define $\eta$ as the fraction of naive T cells in an organism that
is drawn in a sample and assume that all naive T cells in the organism
have the same probability $\eta$ of being sampled. This is true only
if naive T cells carrying different TCRs are not preferentially
partitioned into different tissues and are uniformly distributed
within an animal. Let us assume that a specific TCR is represented by
$\ell$ cells in an organism.  If $N^{*}\eta \gg \ell$, the probability
that $k$ cells from the same clone are sampled approximately follows a
binomial distribution with parameters $\ell$ and $\eta$
\cite{STUMPF,XU_PLOSCB,LEVINA2017,LYTHE_SAMPLING,LYTHE2018}

\begin{eqnarray}
\P{k\vert\ell}\approx 
\binom{\ell}{k}\eta^{k}\left(1-\eta\right)^{\ell-k}, \quad k\leq \ell.
\label{eq:sampling_clone}
\end{eqnarray}
The associated mean \textit{sampled} clone count $c^{\rm s}_{k}$
depends on the clone count $c_{\ell}(\alpha,r\vert \mu^{*})$ predicted
in the body via the formula

\begin{equation}
c^{\rm s}_{k}(\alpha,r\vert \mu^{*},\eta) \approx \sum_{\ell \geq k}c_{\ell}(\alpha,r\vert \mu^{*})
\P{k\vert \ell} = \sum_{\ell \geq k}
c_{\ell}(\alpha,r\vert \mu^{*})\binom{\ell}{k}\eta^{k}(1-\eta)^{\ell-k},
\label{eq:sampling-1}
\end{equation}
where $c_{\ell}(\alpha, r \vert \mu^{*})$ is determined by
Eq.~\ref{eq:ck_alpha_r}. As mentioned, the clone counts
$c_{\ell}(\alpha,r\vert \mu^{*})$ and $c_{k}^{\rm s}(\alpha,r\vert
\mu^{*}, \eta)$, before averaging over $\pi(\alpha,r)$, depend
implicitly on $\lambda/\bar{\alpha}$ and other parameters that define
$\pi_{r}(r)$ through the determination of $\mu^{*}$ via
Eq.~\ref{eq:fixed-point-equation_2}.  The sum in
Eq.~\ref{eq:sampling-1} can be explicitly performed to yield

  \begin{equation}
    c_{k}^{\rm s}(\alpha,r\vert \mu^{*},\eta) = {Q\over k!}
   \left({\eta r/\mu^{*} \over
      1-(1-\eta)(r/\mu^{*})}\right)^{k}\left({1-r/\mu^{*}\over
      1-(1-\eta)(r/\mu^{*})}\right)^{\!{\alpha\over r}}
    \,\prod_{j=0}^{k-1}\left({\alpha\over r}+j\right).
    \label{CKS_ar}
  \end{equation}
The total expected number of clones predicted in
the sample comprised of cells within $[\alpha, \alpha+\dd \alpha]$ and
$[r, r+\dd r]$ can be found from direct summation:

\begin{equation}
  C^{\rm s}(\alpha,r \vert \mu^{*},\eta) =\sum_{k=1}^{\infty}
  c_{k}^{\rm s}(\alpha,r \vert\mu^{*},\eta) = Q\left[1-
\left({1-r/\mu^{*} 
\over 1- (1-\eta)r/\mu^{*}}\right)^{\alpha/r}\right].
\label{CS}
\end{equation}

\begin{figure}
\begin{center}
  \includegraphics[width=5.2in]{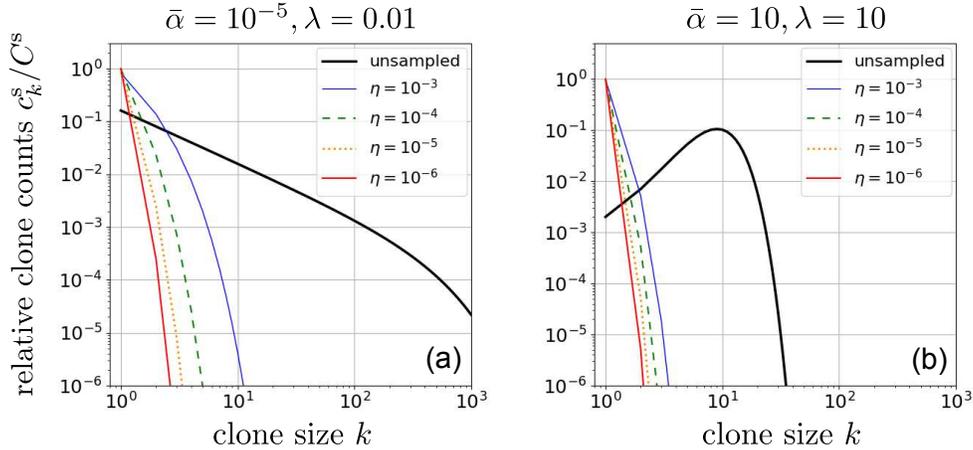}
  \end{center}
\vspace{1mm}
\caption{The effects of sampling on two different
    \text{relative} neutral-model ($r=1/2$) clone counts $c_{k}^{\rm
      s}/C^{\rm s}$. In (a), we used $\bar{\alpha} = 10^{-5},
    \lambda=0.01$ Sampling increases the exponential decay in
    $c_{k}$. Plots of the relative clone counts from the simple
    neutral model (Eqs.~\ref{CKS_ar} and \ref{CS} or Eqs.~\ref{CKS_I}
    and Eq.~\ref{CT_I} in Appendix \ref{FORMS}).  The effect of
    sampling is illustrated for $\eta = 1, 10^{-3}, 10^{-4}, 10^{-5}$,
    and $10^{-6}$. In (b), a very different set of parameters,
    $\bar{\alpha}=\lambda=10$, leads to a qualitatively different
    clone count pattern exhibiting a peak. Even when the predicted
    $c_{k}$ are very different for different parameters, very small
    sampling produces $c_{k}^{\rm s}$ that do not differ
    significantly.
  \label{SAMPLING}}
\end{figure}

The effects of sampling are shown in Fig.~\ref{SAMPLING}. Subsampling
greatly affects the observed clone counts, increasing the exponential
decay of $c_{k}$ \cite{LEVINA2017}. This shifts $c_{k}$ at large $k$
to much smaller values of $k$, while reducing the values of $c_{k}$
for small $k$. In Fig.~\ref{SAMPLING}(a) and (b) we use two very
different parameter values $\alpha=10^{-5}, r=1/2, \lambda = 0.01$ and
$\alpha=\lambda =10, r=1/2$ to generate two qualitatively very
different $c_{k}$. If the subsampling $\eta \ll 1$ is sufficiently
small, the resulting $c_{k}^{\rm s}$ corresponding to qualitatively
different $c_{k}$ can become qualitatively similar. This implies that
sampling may make the estimation of whole-body clone counts from
sampled data somewhat ill-conditioned, \textit{i.e.}, different
whole-body clone counts, upon sampling, may yield similar sampled
clone counts.

Although sampling has a dominating effect on the inference of $c_{k}$,
immigration and proliferation rate distributions may also strongly
affect $c_{k}$. By averaging the clone counts $c_{k}(\alpha,r\vert
\mu^{*})$ (Eq.~\ref{eq:ck_alpha_r}), the sampled clone counts
$c_{k}^{\rm s}(\alpha,r\vert \lambda,\eta)$ (Eq.~\ref{CKS_ar}), and
the total clone count $C^{\rm s}(\alpha,r\vert \mu^{*},\eta)$
(Eq.~\ref{CS}) over the distribution $\pi(\alpha,r)$, we can make
predictions on expected clone counts in the body and in the sample.

\subsection*{Heterogeneity and determination of $\pi(\alpha,r)$}

The fundamental result given in Eq.~\ref{CKS_ar} applies only to the
clone count density (expected clone counts with immigration rate
within $[\alpha, \alpha+\dd \alpha]$ and proliferation rates within
$[r, r+\dd r]$).  Now, we propose forms for $\pi(\alpha,r)$ in order
to evaluate $c_{k}^{\rm s}(\mu^{*}, \eta) = \int_{0}^{\infty}\!\!\dd
\alpha \int_{0}^{1}\!\!\dd r \, \pi(\alpha,r) c_{k}^{\rm s}(\alpha,r
\vert \mu^{*},\eta)$.  For simplicity, we assume that $\alpha$ and $r$
are uncorrelated and that the distribution factorises:
$\pi(\alpha,r)=\pi_{\alpha}(\alpha)\pi_{r}(r)$.

\paragraph*{Proliferation rate heterogeneity:} 
First, we consider a distribution of TCR sequence-dependent
proliferation rates. Since TCR-antigen affinity depends on the
receptor amino-acid sequence, the rate of T cell activation and
subsequent proliferation is clone-specific
\cite{PROLIF_PNAS,MAYER_PNAS_2019}.  Thus, the specific interactions
between TCRs and low-affinity MHC/self-peptide complexes maps to a
distribution of proliferation rates among all the $Q$ possible clones.
Since there are no data (known to us) that can be used to infer this
mapping or the specific shape of $\pi_{r}(r)$, we assume, for
simplicity, a simple uniform ``box'' distribution centered about a
mean value $\bar{r}=1/2$:

\begin{equation}
\pi_{r}(r)=
\begin{cases}
{\displaystyle{1/w}} & \text{if } |r-1/2|<w/2 \\
0 & \text{otherwise},
\end{cases}
\label{eq:pi_r}
\end{equation}
where $0\leq w\leq 1$ represents the relative width of the uniform box
distribution.  The dimensionless self-consistency condition
(Eq.~\ref{eq:fixed-point-equation_2}) thus yields

\begin{equation}
  \mu^{*} = {\left({1\over 2}+{w\over 2}\right)
    e^{\lambda w/\bar{\alpha}}-\left({1\over 2}-{w\over 2}\right)
    \over e^{\lambda w/\bar{\alpha}}-1},
\label{MUSTAR}
\end{equation}
where $\bar{\alpha}$ is the mean per-clone immigration rate normalized
by the maximum proliferation rate $R$.

\begin{figure}
\centering{\includegraphics[width=6.5in]{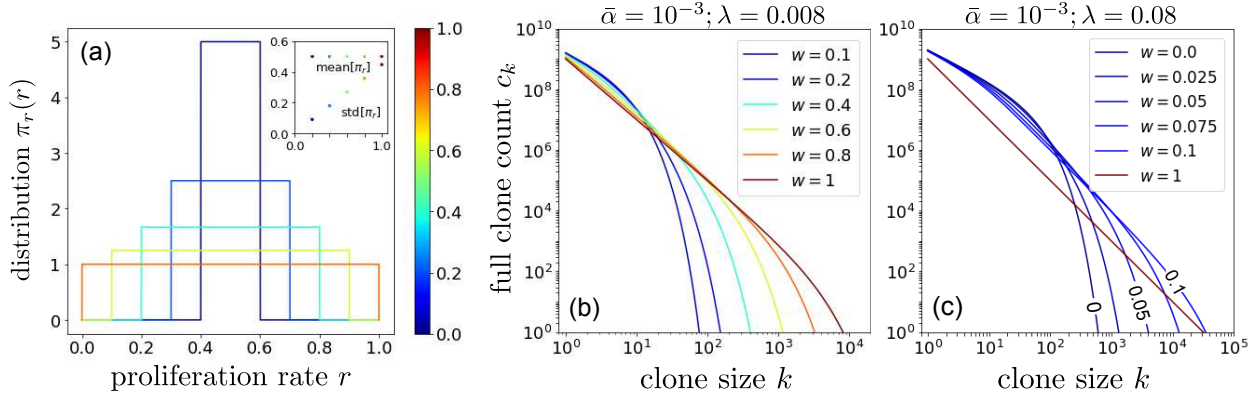}}
\vspace{2mm}
\caption{An exploration of the effects of proliferation
    rate heterogeneity on the mean clone counts $c_{k}$ with
    $Q=10^{13}$ (a) Various box distributions illustrated for
    dimensionless $w=0,0.2,0.4,0.6,0.8$, and $1$.  (b) Using
    Eq.~\ref{MUSTAR} and the dimensionless values
    $\bar{\alpha}=10^{-3}, \lambda = 8\times 10^{-3}$ such that
    $\lambda/\bar{\alpha} = 8$, we plot the corresponding clone counts
    and show that wider distributions typically generate longer-tailed
    $c_k$. However, if $\lambda$ is set even larger such that
    $\lambda/\bar{\alpha} = 80$, (c) shows that even modest values of
    $w$ can generate a very long-tailed $c_{k}$. In this very
    large-$\lambda/\bar{\alpha}$ case, the effects of heterogeneous
    proliferation saturate at very small $w$ beyond which has
    negligible effect in further extending the tail.
  \label{CK_W}}
\end{figure}

Before we use these formulae to understand the experimentally measured
clone counts, we explore the effects of proliferation rate
heterogeneity on the clone counts.  In Fig.~\ref{CK_W}, we plot the
predicted, whole-organism ($\eta=1$) clone counts for different values
of $w$. Since the function $c_{k}$ has an exponentially decaying term
$(r/\mu^{*})^k$, conditions that lead to larger $\mu^{*}$ lead to the
slowest decaying $c_{k}$.
A fixed normalized value of $r=1/2$ leads to an exponential decay in
$c_{k}$ of approximately $2^{-k}$. However, if $r$ is integrated over
its full width of values $(0,1)$, the contribution from $r\approx 1$
generates a longer-tailed distribution $c_{k}$.  Note that for large
$\lambda/\bar{\alpha} \gtrsim 1$, \textit{smaller} $w$ gives rise to
smaller $\mu^{*}$, and even a modest $w$ is enough to make the $c_{k}$
decay slowly.

\paragraph*{Immigration rate heterogeneity:} Next, we propose a way of evaluating 
formulating a distribution $\pi_{\alpha}(\alpha)$ that represents the
TCR sequence-dependent output from the thymus.  A specific form for
$\pi_{\alpha}(\alpha)$ can be obtained from previous studies that
predict V(D)J recombination frequencies associated with each TCR
sequence. The statistical model for differential V(D)J recombination
in humans is implemented in the Optimized Likelihood estimate of
immunoGlobulin Amino-acid sequences (OLGA) software \cite{OLGA}, which
is an updated version of the Inference and Generation of Repertoires
(IGoR) software \cite{Marcou2018}. Below, we estimate
$\pi_{\alpha}(\alpha)$ by sampling a large number of TCRs from OLGA
that draws sequences according to their generation probability.  Our
working assumption is that thymic selection is uncorrelated with V(D)J
recombination so the relative probabilities of forming different TCRs
provide an accurate representation of the ratios of the TCRs exported
into the periphery.



Both IGoR and OLGA can be used to generate the probabilities
corresponding to each drawn sequence but this requires significant
computational time and memory. However, since the sequence draws are
proportional to the underlying probabilities, we simply drew
$N_{\star}=10^9$ sequences and exploited the probability-biased
selection of sequences by counting the frequencies of each
sequence. Out of $N_{\star}$ sequence draws from IGoR or OLGA, there
will be $C_{\star}$ distinct sequences with $b_{j}$ of them occurring
$j$ times.

%
%

\begin{figure}
\begin{center}
\includegraphics[width=5.4in]{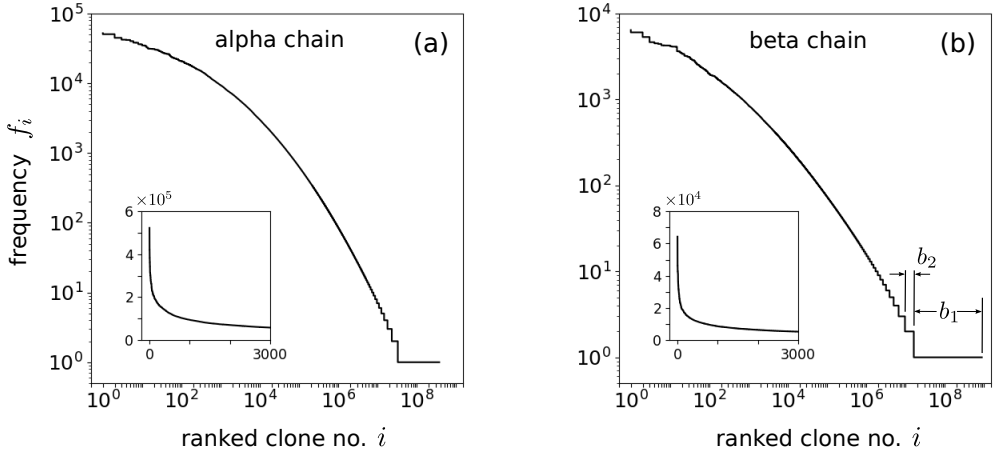}
\end{center}
\caption{Ordered integer-valued frequencies $f_{j}$ of
  $N_{\star}=10^{9}$ TCR sequences drawn from (a) alpha, and (b) beta
  chain sequence draws using OLGA, plotted on a log-log
  scale. $C_{\star}< N_{\star}$ is the total number distinct sequences
  drawn and $b_{j}$ is the number sequences that exhibit the specific
  frequency $f_{j}$. Since $C_{\star}$ is comparable to $N_{\star}$,
  the drawn sequences are dominated by those of low probability that
  appear only once. The insets display the frequencies on a linear
  scale and indicate the long tail behavior of the frequencies. The
  shape of the frequency spectra is self-similar once $N_{\star}
  \gtrsim 10^{7}$, allowing us to use this sampling procedure to
  reliably estimate $\pi_{\alpha}(\alpha)$. \label{OLGA_9}}
\end{figure}
For the alpha and beta chains, we found the frequencies $f_{j}$, $j =
1,2,3,\ldots, J$, where $J$ is the total number of different
integer-valued frequencies observed.  The frequencies of the drawn
clones are plotted in Figs.~\ref{OLGA_9}(a) and (b) in decreasing
order.  For the alpha chain, we found $C_{\star}=372,806,648 \approx
3.72\times 10^{8}, J=52,294$ while for the beta chain we found
$C_{\star}=875,920,705 \approx 8.76\times 10^{8}, J=6430$.  The
probability that a sequence has integer-valued frequency $f_j$ is
$b_{j}/C_{\star}$. From these numbers, we can effectively average over
$\pi_{\alpha}(\alpha)$ by taking a sum over discrete values of
$\alpha$: $\int \pi_{\alpha}(\alpha) c_{k}^{\rm s}(\alpha)\dd \alpha
\to \sum_{j=1}^{J}{b_{j}\over C_{\star}} c_{k}^{\rm s}(\alpha_{j})$
with

\begin{equation}
\alpha_{j} = {\bar{\alpha} f_{j} \over (N_{\star}/C_{\star})}, \quad
j=1,2,\ldots J,
\label{ALPHAJ}
\end{equation}
where here, $\bar{\alpha}$ is the dimensionless per-clone immigration
rate, averaged over the set of $C_{\star}$ unique sequences.

Now that we have defined the possible distributions for
$\pi_{\alpha}(\alpha)$ and $\pi_{r}(r)$, we can compute the mean,
sampled, immigration- and proliferation-averaged clone counts and
compare them with measurements. The full formula for the clone counts
is thus

\begin{align}
  c_{k}(\bar{\alpha}, \lambda, w, \eta) & = \int_{0}^{\infty}\!\!\dd \alpha
  \int_{0}^{R}\!\!\dd r \, \pi_{\alpha}(\alpha)\pi_{r}(r)\, c_{k}^{\rm s}(\alpha,r \vert \mu^{*},\eta) \nonumber \\
  \: & = {Q \over k!}\sum_{j=1}^{J}{b_{j}\over C_{\star}}
  \int_{(1-w)/2}^{(1+w)/2}\!{\dd r\over w}\left({\eta r/\mu^{*}
    \over 1-(1-\eta)r/\mu^{*}}\right)^{k}
  \left({1-r/\mu^{*}\over 1-(1-\eta)r/\mu^{*}}\right)^{\!{\alpha_{j}\over r}}\,\prod_{j=0}^{k-1}\left({\alpha_{j} \over r}+j\right),
  \label{CKS_IV}
\end{align}
where $\alpha_{j}(\bar{\alpha})$ is given by Eq.~\ref{ALPHAJ} and
$\mu^{*}$ is given by Eq.~\ref{MUSTAR}. Eq.~\ref{CKS_IV} is thus our
``full model'' from which we make predictions of clones count-related
quantities and compare them with data.  Using this equation, we can
mathematically study the importance of the heterogeneities in $\alpha$
and $r$ by comparing predictions from simpler forms of
$\pi_{\alpha}(\alpha), \pi_{r}(r)$.  In Appendix \ref{FORMS} in the
SI, we provide explicit dimensionless formulae for different
combinations of $\pi_{\alpha}(\alpha)$ and $\pi_{r}(r)$. For example,
we can consider fixed $\alpha$ or $r$ by using $\pi_{\alpha}(\alpha) =
\delta(\alpha-\bar{\alpha})$ or $\pi_{r}(r) = \delta(r - {1\over 2})$,
respectively.

The effective distribution $b_{j}$ over different values $\alpha_{j}$
indicates that most drawn clones only appear once since $b_{1}\gg
b_{j\geq 2}$. Thus, we hypothesize that for sufficiently small
$\bar{\alpha}$, our formulae for $c_{k}$ and all subsequent quantities
can be simplified by taking the limit $\bar{\alpha}/r \ll 1$. Indeed,
as we show in Appendix~\ref{app:small_alpha}, such a simpler
expression remains highly accurate provided the dimensionless
$\bar{\alpha} < 10^{-2}$ and allows efficient computation. Moreover,
this implies that, mathematically, the result arising from
weighting $c_{k}(\alpha,r)$ over $\pi_{\alpha}(\alpha)$ can be
approximated by a single effective value $\bar{\alpha}$, reinforcing
our overall conclusion that proliferation rate heterogeneity is the
dominant factor. Nonetheless, we will use the full summation over
$\alpha_{j}$ (Eq.~\ref{CKS_IV}) in our subsequent analyses.

\section*{Results and Analysis}

Before performing a quantitative comparison with measured clone counts
from Oakes \textit{et al.} \cite{Oakes2017}, we discuss the
qualitative features of our model and typical physiological parameter
ranges. While even the basic model parameters are difficult to measure
and vary widely across different organisms and individuals, our
nondimensionalized model unifies the mechanisms and concepts common to
the maintenance of diversity in the T cell repertoire across different
organisms.

When considering the data, we observe that even after significant
subsampling, there are appreciable clone counts at reasonably large
clone sizes $k$, whereas the unsampled clone counts decay
exponentially in $k$ with rate $\log (\mu^{*}/r)$. Even though $r$ may
take on a range of values, as determined by $\pi_{r}(r)$, the slowest
decay of $c_{k}$ in $k$ arises from the largest possible values of
$r$.  Thus, a larger proliferation rate heterogeneity $w$ will
generally yield a longer-tailed $c_{k}$, as illustrated in
Fig.~\ref{CK_W}.
  
%

Since the data we will analyze are derived from human samples, we will
use the following arguments as a guide to roughly estimate the
relevant range of parameters:

\begin{itemize}
    \item The average total number of \textit{statistically relevant
      circulating} naive T cells is not completely known but estimates
      are $N^{*}\sim 10^{11}$ \cite{Jenkins2010}.  However, the
      circulating population in the peripheral blood is approximately
      two orders of magnitude smaller. These circulating naive T cells
      nonetheless exchange with those in the much larger population in
      the lymph and other tissues. The timescale of this exchange
      (relative to the age of the organism being sampled or the
      intersample times) will determine the effective $N^{*}$ relevant
      for sampling clone counts $c_{k}^{\rm s}$. Thus, we will use an
      order-of-magnitude estimate on the lower range of measurements
      and estimate $N^{*}\sim 10^{10}-10^{11}$.
    \item The total possible number $Q$ of TCRs of either alpha or
      beta chains may be in the range $10^{13}-10^{18}$
      \cite{Davis1988}, but the effective number is probably much
      smaller as many clones may never be generated in a
      lifetime. Thus, an \textit{effective} value of $Q$ may reside at
      the lower range which leads to $\lambda\equiv N^{*}/Q \sim
      10^{-4}-10^{-2}$.
    \item The average (dimensional) immigration rate per clone $\bar{\alpha}$ can be
      deduced from the total thymic output of all clones
      $\bar{\alpha}Q$, which has been estimated across a wide range of
      values $\bar{\alpha}Q \sim 10^{7}-10^{8}$/day
      \cite{TCELLINHIV,TREC_KIRSCHNER,TREC,YATES2009,Westera2015}. If
      we use an effective repertoire size of $Q \sim
      10^{13}-10^{14}$, the average per clone immigration rate becomes
      $\bar{\alpha} \sim 10^{-7}-10^{-5}/\text{day}$.
    \item The mean proliferation rate $\bar{r}$ is difficult to
      measure but has been estimated to be on the order of $\bar{r}
      \sim 10^{-4}-10^{-3}/\text{day}$ \cite{Westera2015}. If we
      nondimensionalize using $2\bar{r}$, the \textit{dimensionless}
      $\bar{\alpha} \sim 10^{-4}-10^{-1}$.
    \item The sampling fraction $\eta$, although in principle
      determined experimentally, is also hard to quantify due to
      the uncertainty in $N^{*}$. Blood sampling \textit{volume}
      fractions from humans are typically $\eta\sim 10^{-3}$; however,
      in recent experiments \cite{Oakes2017} the number of enumerated
      sequences $\sim 10^5$, which, given rough estimates of effective
      $N^{*}\sim 10^{10}-10^{11}$, yield $\eta \sim 10^{-6}-
      10^{-4}$. Due to this uncertainty in $\eta$, we will explore
      different fixed values of $\eta$ around $10^{-5}$.
\end{itemize}


Using the above guide for reasonable parameter ranges, we now consider
fitting our results in Eqs.~\ref{CKS_IV}, \ref{CKS_I}, \ref{CT_I},
\ref{CKS_II}, \ref{CT_II}, \ref{CKS_III}, and \ref{CT_III} to some of
the available data \cite{Oakes2017}.  Before doing so, note that
although the log-log plots shown in Figs.~\ref{DATA}(a,b) provide a
simple visual for $\log c_{k}^{\rm s}$ or $\log[c_{k}^{\rm s}/C^{\rm
    s}]$, fitting must be performed on the linear scale.

The measured data includes data at values of $k$ for which no clones
were detected so that $c_{k}^{\rm s} = 0$. On the log scale these data
points include $\log c_{k}^{\rm s} \to -\infty$ which nonetheless
should be included in the fitting as they represent realizations of
the system.  Numerical fitting on the log-log scale would be
misleading once a value of $c_{k}^{\rm s} = 0$ is encountered. Thus,
will fit our mean-field model on the linear scale to the
\textit{fraction} $f_{k}^{\rm s}$ of the total number of sampled cells
that are in clones of size $k$:

\begin{equation}
f_{k}^{\rm s}(\bar{\alpha},\lambda, w, \eta) \equiv
{kc_{k}^{\rm s}(\bar{\alpha},\lambda, w, \eta) \over N^{\rm s}}=
{kc_{k}^{\rm s}(\bar{\alpha},\lambda, w, \eta) \over 
\sum_{\ell=1}^{\infty}\ell c_{\ell}^{\rm s}(\bar{\alpha},\lambda, w, \eta)} =
{kc_{k}^{\rm s}(\bar{\alpha},\lambda, w, \eta) \over Q \eta \lambda},
\label{FKS}
\end{equation}
where the denominator $Q \eta \lambda$ comes directly from the
definition $\sum_{\ell=1}^{\infty}\ell c_{\ell}^{\rm
  s}(\bar{\alpha},\lambda\vert\eta)\equiv N^{\rm s}$, the sampling
relation $N^{\rm s} = \eta N^{*}$, and
Eq.~\ref{eq:fixed-point-equation_2}. Rather than using $N^{\rm s}$
directly from the number of reads in an experimental sample,
equivalently, we use the model expression $N^{\rm s} = Q \eta\lambda$
to arrive at the last equality in Eq.~\ref{FKS}. This form ensures
strict normalization $\sum_{k=1}^{\infty}f_{k}^{\rm s} = 1$ and is
independent of the unknown repertoire size $Q$ ($c_{k}^{\rm s}$ is
proportional to $Q$). Note that for $f_{k}^{\rm s}\equiv kc_{k}^{\rm
  s}/N^{\rm s}$, the implicit factor of $Q$ in $c_{k}^{\rm s}$
(Eq.~\ref{CKS_ar}) can cells the explicit $Q$ in the denominator of
Eq.~\ref{FKS}. Thus, $f_{k}^{\rm s}$ as well as $c_{k}^{\rm s}/C^{\rm
  s}$ depend on the effective value of $Q$ only through the
determination of $\mu^{*}$ through $\lambda\equiv N^{*}/Q$ in
Eq.~\ref{eq:fixed-point-equation_2}.


Our mathematical framework provides only \textit{mean} sampled clone
counts while each sample of the data represents one realization.
Large sample-to-sample variations in the clone counts would render the
fitting less informative, but these large variations were not seen in
the triplicate samples in Oakes \textit{et al.}
\cite{Oakes2017}. Mechanistically, we expect that for large $k$ the
number of cells contributing to $f_{k}^{\rm s}$ is also large so
demographic stochasticity is relatively small and results in small
uncertainties in the value of $k$, and not in the magnitude of
$f_k^{\rm s}$.  Large clones are also likely include memory T cells
that have been produced after antigen stimulation of specific
clones. Memory T cells are difficult to accurately distinguish from
naive T cells \cite{Oakes2017} but we will see that large $k$
components of $f_{k}^{\rm s}$ negligibly influence the fitting.

For small $k$, the numbers of clones are large so we expect sampling
and intrinsic stochasticity to also be controlled. Thus, we averaged
the clone counts from three different blood samples from each patient
in Oakes \textit{et al.} \cite{Oakes2017} and only compared them to
the expected value of the sampled clone counts in our model.

By comparing our model $f_{k}^{\rm s}(\bar{\alpha},\lambda, w, \eta)$
with the data $f_{k}^{\rm s}({\rm data})$ via the total least-squares
error

\begin{equation}
H(\bar{\alpha}, \lambda, w,\eta) = \sum_{k=1}^{\infty}\vert f_{k}^{\rm
  s}({\rm data}) - f_{k}^{\rm s}(\bar{\alpha}, \lambda, w,
\eta)\vert^{2},
\end{equation}
we explore how $H(\bar{\alpha}, \lambda, w,\eta)$ depends on the
parameters $\bar{\alpha}$, $\lambda$, $w$, and sampling fraction
$\eta$.

Since the variation was not large and there were already too many
uncertainties (particularly the sampling fraction $\eta$, the shape of
$\pi_{r}(r)$, and the accuracy of a steady-state approximation), we
did not pursue a much more technical uncertainty analysis.  Our goal
was to simply consider the data using a reasonable initial model that
captures the known general features of naive T cell maintenance
(immigration, birth, death) and find constraints among the parameters
$\lambda, \bar{\alpha}$, and $w$.

\begin{figure}[h]
\begin{center}
  \includegraphics[width=6in]{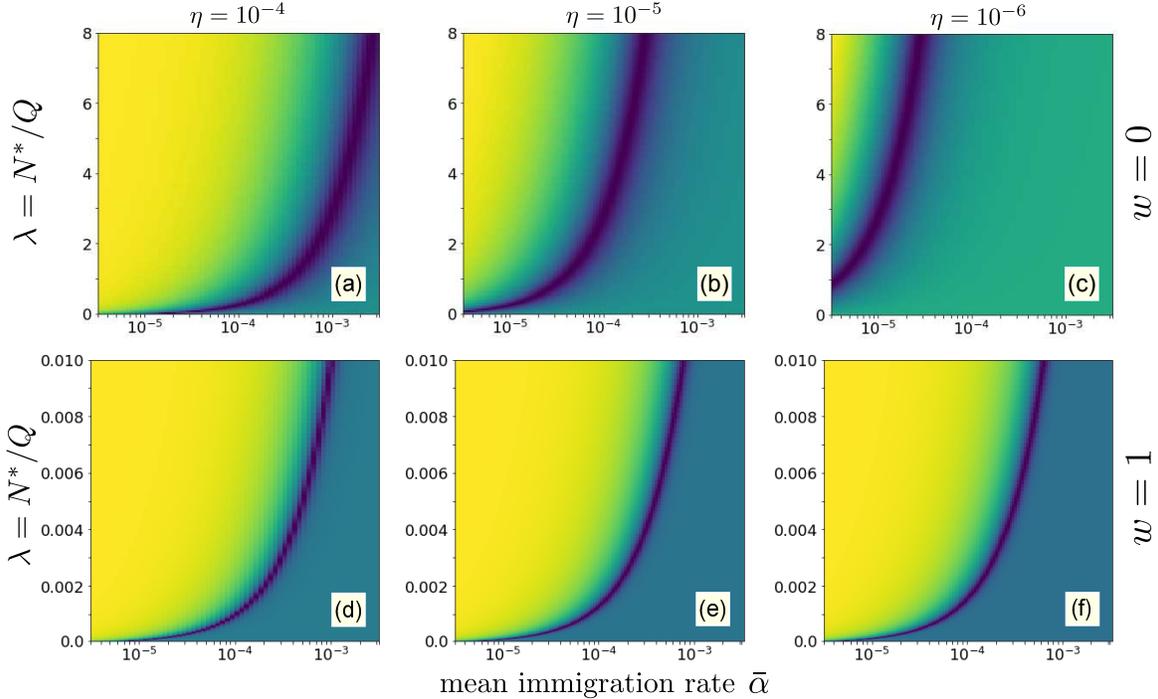}
  \end{center}
\vspace{2mm}
\caption{\label{ERROR_II} The error $H(\bar{\alpha}, \lambda, \eta)$
  plotted as a function of dimensionless $\bar{\alpha}$ (on a log
  scale) and $\lambda$. Darker colors represent smaller values of
  error. The data used are the clone counts of beta chain sequences of
  naive CD4 cells from one patient, averaged over three
  samples. Panels (a-c) uses the simple neutral model
  (Eqs.~\ref{CKS_I} and \ref{CT_I}) sampling fractions $\eta =
  10^{-4}, 10^{-5}$, and $10^{-6}$, respectively. Note that the error
  is minimal along an upward path relating $\lambda$ to
  $\bar{\alpha}$. For the neutral model, the error is very sensitive
  to the sampling fraction $\eta$.  For a fixed, physiologically
  reasonable value of $\bar{\alpha}$, the optimal $\lambda$ is
  unreasonably too large ($>1$). The error along the valley does not
  change appreciably and only slightly decreases as $\lambda$ and
  $\bar{\alpha}$ become smaller.  In the full-width ($w=1$)
  distributed proliferation rate model shown in (d-f), the errors are
  insensitive to $\eta \sim 10^{-6}- 10^{-4}$ and the optimal
  $\lambda=N^{*}/Q$ values are much smaller, consistent with our
  estimates of $N^{*}$ and repertoire size.}
\end{figure}

In Figs.~\ref{ERROR_II}(a-c) the data $f_{k}^{\rm s}(\rm data)$ were
derived from the average of three measurements from beta chain CD4
sequences sampled from one patient \cite{Oakes2017}.  Using this data,
we compute and plot the error $H(\bar{\alpha},\lambda, w=0,
\eta=10^{-4}, 10^{-5}, 10^{-6}))$ as a function of $\lambda$ for
various values of $\bar{\alpha}$ using the neutral model ($w=0$,
Eq.~\ref{CKS_I} in Appendix \ref{FORMS}).  For reasonable values of
$\bar{\alpha}$ and sampling fraction $\eta$, we see that the predicted
$\lambda \equiv N^{*}/Q$ is typically ${\cal O}(1)$ or larger. In
Figs.~\ref{ERROR_II}(d-f) we use the full-width ($w=1$) distribution
of $\pi_{r}(r)$ to show the analogous error for the same data using
the same sampling fractions $\eta=10^{-4}, 10^{-5}, 10^{-6}$. Note
that the optimal (low error) values of $\lambda$ are significantly
smaller that those in found using $w=0$ (Figs.~\ref{ERROR_II}(a-c))
and that the results are rather insensitive to sampling fraction
$\eta$. These smaller values of $\lambda = N^{*}/Q$ are more
consistent with known physiological understanding. Thus, the
distributed proliferation rate model provides a much more
self-consistent fit to the data than the fixed proliferation rate
neutral model.

\begin{figure}[h]
\centering{\includegraphics[width=6.75in]{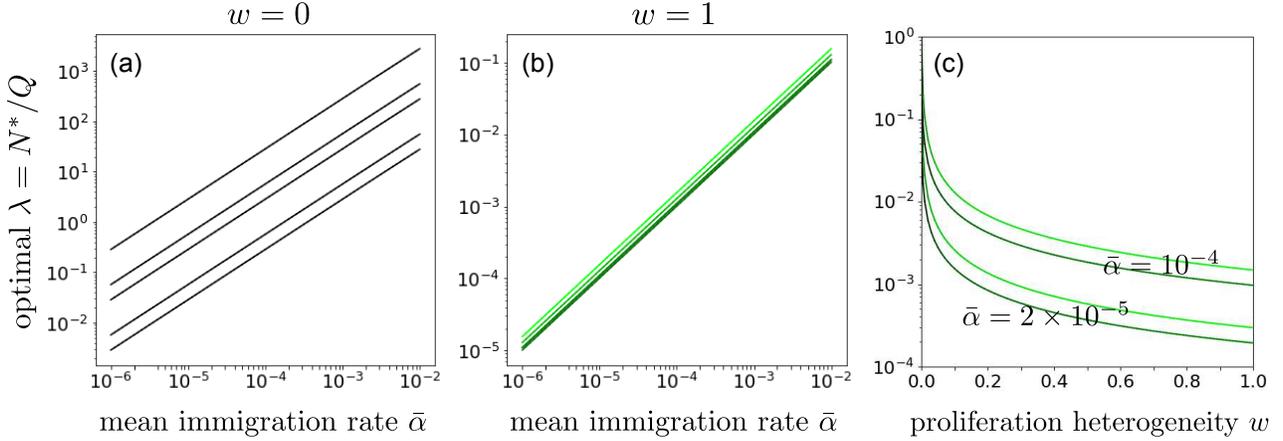}}
\vspace{2mm}
\caption{Log-log plots of optimal $\lambda$ values for fixed
  $\bar{\alpha}$ for (a) the neutral model, and (b) the full-width
  distributed proliferation rate model. These curves trace the values
  of $\lambda$ along the minimum valleys in Fig.~\ref{ERROR_II} and
  show the relative insensitivity of the distributed proliferation
  rate model to sampling fraction. The variation in color of the
  curves in (a) indicates variation in the total error along the
  minimum valley. The error values are nearly constant in the case of
  (b). Both slopes are near one, indicating the optimal $\lambda$ is
  approximately proportional to $\bar{\alpha}$ over a wide range of
  values. Thus, the model and associated data show that an
  ``independent'' parameter is $\lambda/\bar{\alpha}$, the key
  parameter in determining $\mu^{*}$. (c) Log-linear plot of optimal
  $\lambda$ values as a function of proliferation rate heterogeneity
  $w$.  When $\bar{\alpha}$ is fixed ($2\times 10^{-5}$ and $10^{-4}$
  in this example), we find that even a small heterogeneity $w$
  quickly reduces the predicted value of $\lambda$ to below one;
  however, if $\lambda$ is forced to be even smaller, the required
  heterogeneity $w$ increases. The lower darker curves correspond to
  $\eta=10^{-4}$ while the lighter curves correspond to
  $\eta=10^{-6}$. \label{SIGMA_ALPHA}}
\end{figure}

Note that the value of the errors along the minimum valley are nearly
constant, only slightly decreasing as $\lambda, \bar{\alpha}\to
0$. This means that the model and associated data constrains $\lambda$
to $\bar{\alpha}$, but cannot independently determine them. The
constraint arises as a linear relationship as shown in
Fig.~\ref{SIGMA_ALPHA} that shows the relationship between $\lambda$
and $\bar{\alpha}$ along the minimizer of $H(\bar{\alpha},\lambda,
\eta)$.
Figs.~\ref{SIGMA_ALPHA} show the optimal $\lambda$ as a function of
$\bar{\alpha}$ for (a) the neutral model and (b) the full-width
$\pi_{r}(r)$ model. Although the variation in error is negligible
across $\bar{\alpha}$ for each model $w$ (color variation along the
lines), the full-width distributed model ($w=1$) generates a
\textit{larger} error than the neutral model, but it provides much
more reasonable values of $\lambda, \bar{\alpha}$ and is much less
sensitive to $\eta$. To investigate intermediate values of $0<w<1$ to
determine if a low error can be achieved at small values of $\lambda
\ll 1$, we plot in Fig.~\ref{SIGMA_ALPHA}(c) the optimal $\lambda$ as
a function of proliferation rate heterogeneity $w$. Note that even a
small heterogeneity significantly reduces the predicted
$\lambda$. However, if $\lambda\approx 10^{-3}$ or $10^{4}$ say, the
required $w$ can become quite large.

To explore the dependence of the error on the proliferation rate
heterogeneity, we fix $\bar{\alpha}, \lambda$, and $\eta$, and plot
the error $H(\bar{\alpha}, \lambda, w,\eta)$ as a function of $w$.
\begin{figure}[h]
  \begin{center}
\includegraphics[width=6.75in]{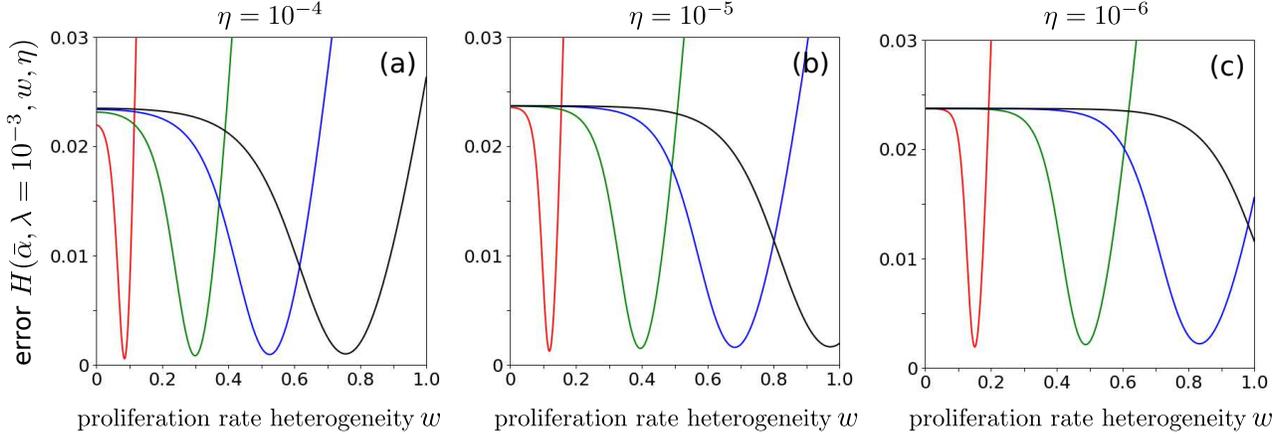}
    \end{center}
\vspace{2mm}
\caption{The error $H(\bar{\alpha},\lambda, w, \eta)$ using CD4 alpha
  data from Oakes et al.  \cite{Oakes2017} plotted as a function of
  $w$ for various $\lambda/\bar{\alpha}$. We fixed $\lambda = 10^{-3}$
  and varied, from left to right, $\bar{\alpha}= 2\times 10^{-5}$
  (red), $6\times 10^{-5}$ (green), $10^{-4}$ (blue) and $1.4\times
  10^{-4}$ (black). From (a) to (c), $\eta = 10^{-4}, 10^{-5}$, and
  $10^{-6}$. Smaller values of $\lambda/\bar{\alpha}$ result in larger
  best-fit values of $w$.
\label{ERROR_W}}
\end{figure}
Fig.~\ref{ERROR_W} shows how the minimizing value of $w$ is sensitive
to $\lambda/\bar{\alpha}$, suggesting larger proliferation
heterogeneity as $\lambda/\bar{\alpha}$ is decreased.  The minimum of
the error, however, is rather insensitive to
$\lambda/\bar{\alpha}$. For sufficiently small $\lambda/\bar{\alpha}$,
the minimum error at $w=1$ is higher than those with minimum $w
<1$. Since the minima in the interior $0<w<1$ are nearly invariant to
$\lambda/\bar{\alpha}$, we indeed show that near-optimal solutions
with $\lambda \ll 1$ can be found when the proliferation rate
heterogeneity $w$ is appreciable.

\begin{figure}[h]
  \begin{center}
\includegraphics[width=5.2in]{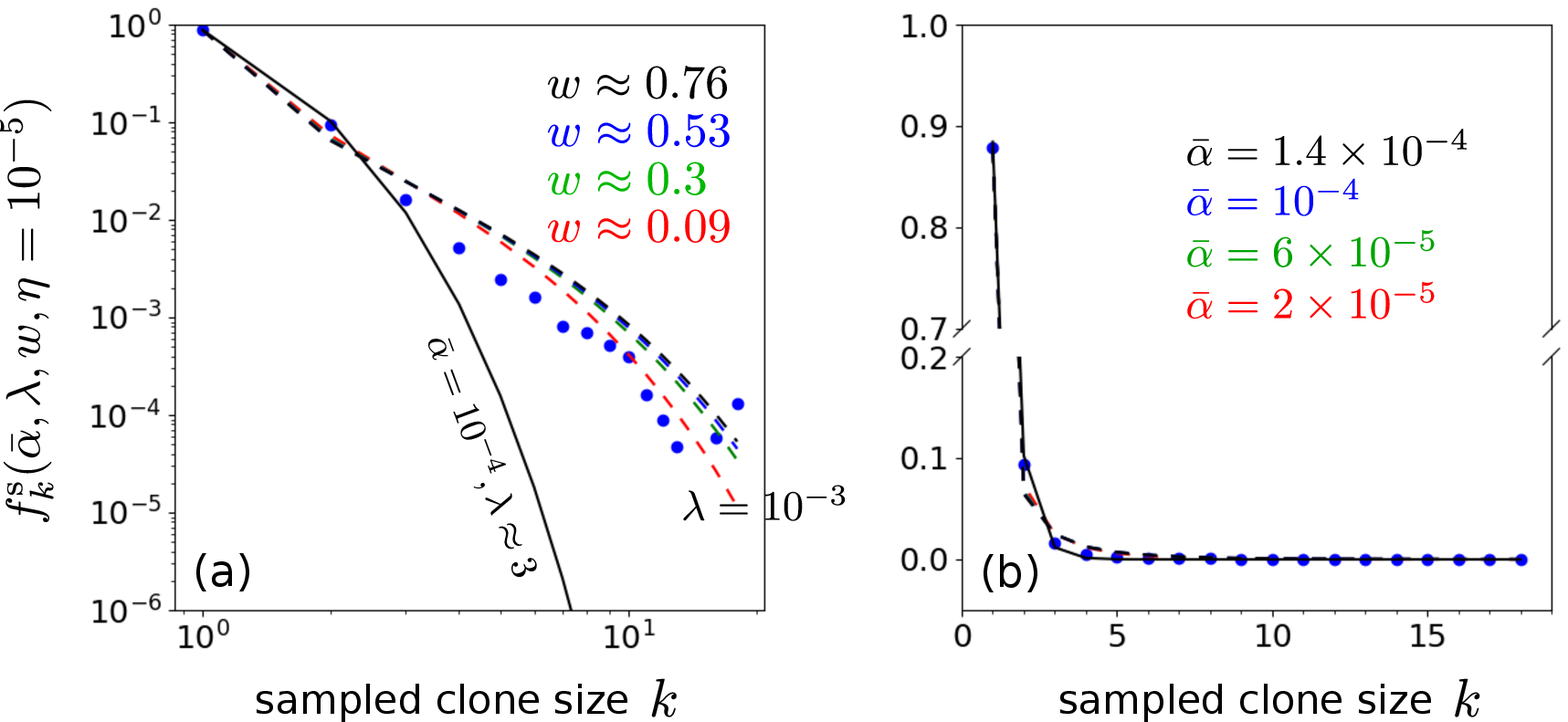}
    \end{center}
\vspace{2mm}
\caption{Plots of the representative optimal solutions of clone counts
  plotted against data.  The model predictions and CD4 beta chain data
  are plotted both in (a) log-log and (b) linear scales (there are no
  zeros in this dataset). In (a), the best fit model for the neutral
  model ($w=0$ and $\pi_{\alpha}(\alpha)=\delta(\alpha-\bar{\alpha})$)
  using $\bar{\alpha}=10^{-4}$ is given by $\lambda \approx 3$ shown
  by the solid black curve. The dashed curves represents best-fit
  curves using the values associated with the error minima in
  Fig.~\ref{ERROR_W}(b), where $\bar{\alpha}= 2\times 10^{-5}$,
  $w\approx 0.09$ (red), $6\times 10^{-5}$,$w\approx 0.3$ (green),
  $10^{-4}$, $w\approx 0.53$ (blue) and $1.4\times 10^{-4}$, $w\approx
  0.76$ (black). Note that the neutral model fits well for only the
  first 2-3 $k$-points, while the heterogeneous model ($w>0$) fits
  better at larger $k$. \label{FK_FIT}}
\end{figure}

Using the parameters associated with the minima in Fig.~\ref{ERROR_W},
we plot our predicted $f_{k}^{\rm s}$ against the data $f_{k}^{\rm
  s}({\rm data})$ in Fig.~\ref{FK_FIT}.
Although the neutral model appears to fit better, especially for small
$k$, the necessary values of $\lambda$ and $\bar{\alpha}$ are too
large and small, respectively. Conversely, when proliferation rate
heterogeneity is allowed, the best-fits have small error and are found
using $\lambda \ll 1$. Note that most of the information in the data
lies in how $f_{k}^{\rm s}({\rm data})$ decreases over the first few
values of $k$. The goodness of fit of our model to the data depends
mostly on the predicted initial decreases in $f_{k}^{\rm
  s}(\bar{\alpha}, \lambda, w, \eta)$.

\section*{Discussion}

Here, we review and justify a number of critical biological
assumptions and mathematical approximations used in our analysis. The
effects of relaxing our approximations are also discussed.

\paragraph*{Factorisation of $\pi(\alpha,r)$.} For mathematical tractability, we 
have assumed $\pi(\alpha,r) = \pi_{\alpha}(\alpha)\pi_{r}(r)$. Given
the typical physiological values of $\bar{\alpha}$, the clone count
formulae derived from our model can be accurately approximated by a
single value of $\bar{\alpha}$. Thus, we expect that the immigration
rate distribution can be approximated by $\pi_{\alpha}(\alpha) =
\delta(\alpha-\bar{\alpha})$.  This allows further approximation of
our formulae as shown in Appendix \ref{app:small_alpha}.  In Appendix
\ref{app:correlated}, we explicitly show that factorisation is an
accurate approximation.

We have also assumed that selection is uncorrelated with the
generation probabilities of the TCR nucleotide sequences encoded in
IGoR/OLGA. The assumption is that the recombination statistics are
uncorrelated with the statistics of thymic selection, a process that
is based on TCR amino acid sequences. However, we note that it has
been suggested that selection pressure may induce a correlation
between TCRs generated and selected \cite{SELECTION2014}. The
corresponding statistics of the frequencies of \textit{selected} TCRs
would be modified from those of the \textit{generated} TCRs shown in
Figs.~\ref{OLGA_9}.  Nonetheless, we assume that the resulting
distribution can still be approximated by a single-$\alpha$ model
which will not qualitatively alter our conclusions.

\paragraph*{Mean-field approximation.} Our mean-field approximation
for the mean clone count $c_{k}$ is embodied in Eq.~\ref{CK_ODE}, where
correlations between fluctuations in the total population $N =
\sum_{k}kc_{k}$ in the regulation term $\mu(N)$ and the explicit
$c_{k}$ terms are neglected.  This approximation has been shown to be
accurate for $k \lesssim N^{*}$ when $\bar{\alpha}Q^{2} > \mu(N^{*})$
\cite{Xu2018a}.  The mean-field results overestimate the clone counts
for $k \gtrsim N^{*}$. Moreover, when the total steady-state T cell
immigration rate is extremely small, the effects of competitive
exclusion dominate and a single large clone arises due to competitive
exclusion \cite{exclusion1960,exclusion1961,Xu2018a}. Nonetheless, an
accurate approximation for the steady-state clone abundance $c_{k}$
can be obtained using a variation of the two-species Moran model as
shown in \cite{Xu2018a}.

For the naive T cell system, because $Q$ is so large, the mean
immigration rate $\bar{\alpha}$ is such that competitive exclusion is
not a dominant feature. Moreover, since $N^{*}\gtrsim 10^{10}$, clones
counts at comparable sizes are not observed and predicted to be
negligible in all models.

The values of $f_{k}^{\rm s}({\rm data})$ become exponentially smaller
for large $k$, our inference is most sensitive to the values of
$f_{k}^{\rm s}({\rm data})$ for small to modest $k$. The information
in the data is primarily manifested by how the $f_{k}^{\rm s}({\rm
  data})$ decays in $k$, we before the mean-field approximation
deviates from the exact solution.  Thus, the parameters associated
with the human adaptive immune system satisfy the conditions for the
mean-field approximation to be accurate, justifying its use in the BDI
model.


%
%
%

\paragraph*{Steady state assumption.} In this study, we only
considered the steady state of our birth-death-immigration model in
Eq.~\ref{eq:ck_alpha_r} because this limit allowed relatively easy
derivations of analytical results. This was also the strategy for
previous modeling work
\cite{GOYAL_BMC,Desponds2016,Lythe2016,deGreef2018,Xu2018a}.  However,
the per-clone immigration and proliferation times may be on the order
of months or years, a time scale over which thymic output diminishes
as an individual ages
\cite{Westera2015,YATES2015,YATES2018,Lewkiewicz2018}.  Indeed, clone
abundance distributions have been shown to show specific patterns as a
function of age \cite{YATES2012,BRITANOVA2016,AGING2018}.

Although $N(t)$, with fixed $\bar{\alpha}$ and $\bar{r}$ relaxes to
steady-state quickly, on a timescale of months, the different
subpopulations of specific sizes described by their number $c_{k}$
relax to quasi-steady steady-state across a spectrum of time scales
depending on the clone sizes $k$ \cite{SL2,Xu2018a}. The timescales of
relaxation of the largest clones can be estimated through the
eigenvalues of linearized the Eqs.~\ref{CK_ODE} to be $\sim 10$ years
\cite{Lewkiewicz2018}. 

%
%
%
%
%
%

In addition to time-dependent changes in $\bar{\alpha}Q$, more subtle
time-inhomogeneities such as changes in proliferation and death rates
have been demonstrated \cite{YATES2015,YATES2018}.  Thus, our
steady-state assumption should be relaxed by incorporation of
time-dependent perturbations to the model parameters $\mu^{*}$ and/or
$\pi(\alpha, r)$.  Longitudinal measurements of clone abundances or
experiments involving time-dependent perturbations would provide
significant insight into the overall dynamics of clone abundances.
Finally, notice that the error slowly decreases as
$\lambda,\bar{\alpha} \to 0$, especially for small $w$. The
requirement that $\lambda \ll 1$ requires a best fit $\bar{\alpha}$
that is extremely small.  At such small $\bar{\alpha}$, the time for
the system to reach steady state scales as $\bar{\alpha}$ and the
validity of the steady-state approximation is nullified.

Thus, given the multiple known underlying time-inhomogeneous processes
that are known to exist, the steady-state may not be strictly reached
and our analysis should be considered as an approximation.

\paragraph*{Clustered Immigration.} Our mean field model assumed that each
immigration event introduced a single naive T cell in the immune
system. However, T cells can divide before leaving the thymus and
reach a homeostatic state in the periphery. This process can be
described by the simultaneous immigration of more than one naive T
cell with the same TCR. Clustered immigration of $q$ cells can be
implemented in the core model for $c_{k}$ (Eq.~\ref{CK_ODE}) via an
immigration term of the form
$\alpha_{q}(c_{k-q}(\alpha_{q},r)-c_{k}(\alpha_{q},r))$, where
$c_{k-q}=0$ for $k-q <0$ (see Appendix \ref{app:clustered}). 
For $q >1$, an informative analytic expression for $c_{k}$ is not
available.  In Fig.~\ref{CK_CLUSTER} in the Appendix, we show the
predicted clone abundance $c_k$ for a neutral model in which
$q=5$. When compared to the case where there is only one cell per
immigration, the clone abundance $c_k$ will have a larger slope for
$k\lesssim q$, making it kink more downward near $k \approx q$. Thus,
from Figs.~\ref{CK_CLUSTER} and ~\ref{FK_FIT}(a), we can see that
paired immigration ($q=2$) would increase $f_{k}^{\rm s}$ for $k=2$,
providing an improved fitting to data over single copy immigration
($q=1$).


Thus, in addition to appreciable sensitivity of the predicted clone
counts to $\pi_{r}(r)$, we also expect clustered immigration defined
through the immigration rates $\alpha_{q}$, $q > 1$ to control the
goodness of fit to data. Indeed, Fig.~\ref{CK_CLUSTER} suggests that
the distribution of immigration cluster sizes $q$, in addition to the
proliferation rate heterogeneity $w$, is an important determinant of
measured clone counts and that $\alpha_{q}$ may be constrained by
data. We leave this for future investigation.



\paragraph*{General conclusions.} We developed a heterogeneous
multispecies birth-death-immigration model and analyzed it in the
context of T cell clonal heterogeneity; the clone abundance
distribution is derived in the mean-field limit.  Unlike previous
studies \cite{Desponds2016}, our modeling approach incorporated
sampling statistics and provided simple formulae, allowing us to
predict clone abundances under different rate distributions for
arbitrarily large systems ($N^{*}\sim 10^{10}-10^{11}$), without the
need for simulation. The properties of the BDI model and the overall
shape of the sampled clone count data renders the first few $k$-values
of $c_{k}^{\rm s}$ or $f_{k}^{\rm s}$ the most important for
determining the constraints among the model parameters. In other
words, only the initial rate of the decrease in $f_{k}^{\rm s}({\rm
  data})$ for small $k$ governs the quality of fitting to the model,
and one should not expect to be able to explicitly infer more than one
or two free parameters.

Our heterogeneous BDI model produced mean sampled clone count
distributions that we could directly compare with measured clone
counts. The unsampled clone counts $c_{k}$ of the neutral model
(homogeneous $\alpha$ and $r$) follow a negative binomial distribution
which is further modified upon sampling and distribution over the
heterogeneous immigration and proliferation rates. Although we
determined $\pi_{\alpha}(\alpha)$ through a tool that used
recombination statistics inferred from cDNA and gDNA sequences
\cite{Marcou2018,OLGA}, we found that the behavior of the model is
rather insensitive to distributions $\pi_{\alpha}(\alpha)$ with mean
values $\bar{\alpha}$ much smaller than the largest proliferation
rates $r$. The model results are dominated by many low
immigration-rate clones and a model that replaces $\alpha$ with its
mean value $\bar{\alpha}$ is sufficient.

Conversely, we find that the shape of the clone count profiles $c_{k}$
are quite sensitive to the proliferation rate heterogeneity $w$. A
small amount of heterogeneity quickly reduces the best-fit values of
$\lambda$ to reasonable values.  For estimated values $\eta \sim
10^{-6} - 10^{-4}$, $\bar{\alpha} \sim 10^{-4}$, and small values of
$\lambda = N^{*}/Q \lesssim 10^{-3}$, requires a best-fit width
$w\approx 1$. Heterogeneity is needed to generate clones of
sufficiently large size that persist after sampling. Although the
number of TCR clones with large proliferation rates $r$ may be small,
such clones proliferate more rapidly contributing to higher clone
counts at larger sizes. In particular, we found that the shape of
expected clone abundance is sensitive to the behavior of the
proliferation rate distribution near the maximum dimensional
proliferation rate $R$, $\pi_{r}(r \approx R)$.
%


The predicted clone counts are also modestly sensitive to the
distribution of immigration cluster sizes $q$ (representing transient
proliferation just before thymic output). When $q>1$ cells of a clone
are simultaneously exported by the thymus, the predicted mean clone
counts decay much more slowly for small $k \lesssim q$ (see
Fig.~\ref{CK_CLUSTER}).  This modification will allow for better
fitting since clustered immigration increases the predicted clone
counts for larger $k$, $c_{2}^{\rm s}, c_{3}^{\rm s}$, etc., and
eventually $f_{2}^{\rm s}, f_{3}^{\rm s}$, etc. Thus, we expect that a
model containing multiple clustered immigration rates $\alpha_{q\geq
  1}$ will lower the error and provide better fitting, particularly at
larger $w$. Additional analysis using a distribution of immigration
cluster sizes may allow this type of clone count data to reveal more
information about the physiological mechanism of naive T cell
maintenance.

Even assuming modest heterogeneity, our work lead to the conclusion
that proliferation heterogeneity is the more important mechanism
driving the emergence of the sampled clone count distributions
$c_{k}^{\rm s}$ (and $f_{k}^{\rm s}$) \cite{Oakes2017}. These results
are consistent with the finding that naive T cells in humans are
maintained by proliferation rather than thymic output
\cite{denBraber2012}.  Since we have only investigated the effects of
a uniform distribution for $\pi_{r}(r)$, further studies using more
complex shapes of $\pi(\alpha, r)$ can be easily explored numerically
using our modeling framework. Different parameter values and rate
distributions appropriate for mice, in which naive T cells are
maintained by thymic output, should also be explored within our
modeling framework.

\section*{Acknowledgements}
This work was supported by grants from the NIH through grant
R01HL146552 and the NSF through grants DMS-1814364 (TC) and
DMS-1814090 (MD). The authors also thank the Collaboratory in
Institute for Quantitative and Computational Biosciences at UCLA for
support to RD.

\bibliography{biblio2}

\clearpage


\setcounter{page}{1}
\renewcommand{\theequation}{S\arabic{equation}}
\setcounter{equation}{0}

\renewcommand{\thefigure}{S\arabic{figure}}
\setcounter{figure}{0}

\section*{How heterogeneous thymic output and 
homeostatic proliferation shape naive T cell receptor clone abundance distributions\\[16pt]
S1: Mathematical Appendices}

\section{Neutral model} 
\label{app:neutral_model}

Here, we review the neutral model to provide insight into the
properties of our heterogeneous BDI model.  When there is no
heterogeneity in either proliferation or immigration rates,
$\pi(\alpha,r)=\delta(\alpha-\bar{\alpha})\delta(r-\bar{r})$.  Upon
inserting this expression for $\pi(\alpha, r)$ in
Eq.~\ref{eq:ck_alpha_r}, we find that the clone
abundance $c_k$ follows a negative binomial distribution
\cite{Dessalles2018}:

\begin{equation}
c_{k}=Q\left(1-\frac{\bar{r}}{\mu (N^*)}\right)^{\bar{\alpha}/\bar{r}}
\left(\frac{\bar{r}}{\mu (N^*)}\right)^{k} \frac{1}{k!}
\prod_{k'=0}^{k-1}\left(\frac{\bar{\alpha}}{\bar{r}}+k'\right).
\label{eq:ck_neutral}
\end{equation}
We can also express $c_k/C$, the clone abundance distribution
normalized by the mean richness $C$ in the body as

\begin{eqnarray}
\frac{c_k}{C}=\frac{c_k}{\sum_{\ell\geq 1}c_\ell}
\end{eqnarray}
which is a negative binomial distribution of parameters
$\bar{\alpha}/\bar{r}$ and $\bar{r}/\mu (N^*)$.  Using
$\bar{\alpha}\approx 1.6\times 10^{-8}/\text{day}$, $\bar{r} \sim
5\times 10^{-4}/\text{day}$, and $\mu(N^*) \approx 6.4\times 10^{-4}$,
we find $\bar{\alpha}/\bar{r}\ll\bar{r}/\mu (N^*)$. In this regime,
$c_k/C$, for $k\geq 1$, can be approximated by a log-series
distribution with parameter $\bar{r}/\mu(N^*)$.

To mathematically show that ${c_k}/{C}$ converges to a log-series
distribution when $\bar{\alpha}/\bar{r}\to 0$, consider a random
variable $X$ that follows a negative binomial distribution of
parameters $\bar{\alpha}/\bar{r}$ and $\bar r/\mu(N^*)$

\begin{equation}
\P{X=k} = \left(1-\frac{\bar{r}}{\mu (N^*)}\right)^{\bar{\alpha}/\bar{r}}
\left(\frac{\bar{r}}{\mu (N^*)}\right)^{k} \frac{1}{k!}
\prod_{\ell=0}^{k-1}\left(\frac{\bar{\alpha}}{\bar{r}}+\ell\right).
\label{eq:distrib_X}
\end{equation}
Note that the probability mass function of $X$ is given by $c_k/Q$ as
can be seen from Eq.~\ref{eq:ck_neutral}, the clone abundance
distribution for all possible $Q$ clones, which includes $c_0$, the
number of all clones that are not represented in the organism.  To
find the clone abundance distribution $c_k/C$, for all the $C$ clones
present in the organism, we must exclude the case $k=0$ by
marginalizing the distribution of $X$ over all $X>0$:

\begin{equation}
\P{X=k\vert X>0} = \frac{\P{X=k}}{\sum_{\ell \geq 1}\P{X=\ell}}
=\frac{c_k/Q}{\sum_{\ell\geq 1}c_\ell/Q}=\frac{c_k}{C}.
\end{equation}
What remains is to show that the distribution of converges to a
log-series distribution of parameter $\bar{r}/\mu(N^*)$ when
$\bar{\alpha}/\bar{r} \to 0$.  Consider the moment generating function
of $X\vert X>0$ given by

\begin{equation}
\E{e^{\xi X}|X>0} = \frac{\E{e^{\xi X}}-\E{e^{\xi X}|X=0}
  \P{X=0}}{\P{X>0}}.
\end{equation}
Since the moment generating function of a negative binomial
distribution $\E{e^{\xi X}}$ is known, and since $\P{X>0}=1-\P{X=0}$
(see Eq.~\ref{eq:distrib_X}), we can write

\begin{equation}
\E{e^{\xi X}|X>0} =
\frac{\left(\frac{1-\bar{r}/\mu(N^*)}{1-e^{\xi}\bar{r}/\mu(N^*)}\right)^{\bar{\alpha}/\bar{r}}-\left(1-\frac{\bar{r}}{\mu(N^*)}\right)^{\bar{\alpha}/\bar{r}}}{1-\left(1-\frac{\bar{r}}{\mu(N^*)}\right)^{\bar{\alpha}/\bar{r}}}.\label{eq:mg}
\end{equation}
For any $x>0$, the limit of $\bar{\alpha}/\bar{r}\to 0$, yields

\begin{align*}
x^{\bar{\alpha}/\bar{r}} & =1+\frac{\bar{\alpha}}{\bar{r}}\log
x+o\left(\frac{\bar{\alpha}}{\bar{r}}\right).
\end{align*}
If we apply this result to Eq.~\ref{eq:mg} for $\E{e^{\xi X}|X>0}$, we find

\begin{align*}
\E{e^{\xi X}|X>0} &
=\frac{1+\frac{\bar{\alpha}}{\bar{r}}\log
\left(\frac{\mu(N^{*})-\bar{r}}{\mu(N^{*})-e^{\xi}\bar{r}}\right)
-\left(1+\frac{\bar{\alpha}}{\bar{r}}\log
\left(1-\frac{\bar{r}}{\mu(N^*)}\right)\right)+
o\left(\frac{\bar{\alpha}}{\bar{r}}\right)}{-\frac{\bar{\alpha}}{\bar{r}}
\log\left(1-\frac{\bar{r}}{\mu(N^*)}\right)
+o\left(\frac{\bar{\alpha}}{\bar{r}}\right)}\\
\: & =\frac{\log\left(1-e^{\xi}\frac{\bar{r}}{\mu(N^*)}\right)}
{\log\left(1-\frac{\bar{r}}{\mu(N^*)}\right)}+o\left(1\right),
\end{align*}
which we recognize as the moment generating function of a log series
distribution of parameter $\bar{r}/\mu (N^*)$. Thus, we finally have

\begin{equation}
\lim_{\bar{\alpha}/\bar{r}\to 0} c_{k} = 
\frac{C}{\log\left(\frac{1}{1-\bar{r}/\mu(N^*)}\right)}
{1\over k}\left({\bar{r}\over \mu(N^*)}\right)^{k}.
\label{eq:logseries}
\end{equation}


\section{Explicit forms using different $\pi_{\alpha}(\alpha)$, $\pi_{r}(r)$}
\label{FORMS}

In the following, we propose four simplifying expressions for the
heterogeneity-averaged clone counts $c_{k}^{\rm s}(\lambda\vert \eta)$
derived from Eq.~\ref{CKS_IV}.

\subsubsection*{Clone-independent Neutral model: $\pi(\alpha,r)
  = \delta(\alpha-\bar{\alpha})\delta(r-\bar{r})$}

First, consider the simplest case where all naive T cells carry the
same immigration and proliferation rates $\bar{\alpha}$ and $\bar{r}$,
respectively, and define $\pi(\alpha,r) =
\delta(\alpha-\bar{\alpha})\delta(r-\bar{r})$.  This case corresponds
to $w\to 0$ and $r\to \bar{r} = 1/2$. The self-consistent condition
for $\mu^{*}$ and $\bar{\alpha}/\bar{r}$ become

\begin{equation}
  {\bar{r} \over \mu^{*}}
  \to {\lambda \over \lambda + 2\bar{\alpha}}, \quad  {\bar{\alpha}\over \bar{r}}\to 2\bar{\alpha},
    \label{NEUTRAL_DIMENSION}
\end{equation}
and the clone count can be explicitly simplified to

\begin{equation}
    c_{k}^{\rm s}(\bar{\alpha},\lambda, \eta)
    \equiv {2\bar{\alpha} Q\over k!}  \left({\eta\lambda \over
      \eta\lambda + 2\bar{\alpha}}\right)^{\!k}\left({2\bar{\alpha} \over
      \eta\lambda + 2\bar{\alpha}}\right)^{\!\!2\bar{\alpha}}\,\, \prod_{j=1}^{k-1}(2\bar{\alpha}+j).
\label{CKS_I}
\end{equation}
The total sampled clone count is then

\begin{equation}
  C^{\rm s}(\bar{\alpha},\lambda, \eta) = \sum_{k=1}^{\infty}
  c_{k}^{\rm s}(\bar{\alpha}, \lambda, \eta) =
  Q\left[1-\left({2\bar{\alpha}\over \eta\lambda +
      2\bar{\alpha}}\right)^{2\bar{\alpha}}\right]
  \label{CT_I}
\end{equation}

\subsubsection*{Fixed $\alpha$, distributed $\pi_{r}(r)$:
  $\pi(\alpha,r) = \delta(\alpha-\bar{\alpha})\pi_{r}(r)$}

Next, consider a common immigration rate $\bar{\alpha}$ for all T cell
clones and a box distribution $\pi_{r}$ of full width $w=R=1$.  The
integral over the dimensionless proliferation rate $r$ is now over
$(0,1)$ and $\mu^{*} = (1-e^{-\lambda/\bar{\alpha}})^{-1}$. The
averaged clone counts are now explicitly

\begin{equation}
  c_{k}^{\rm s}(\bar{\alpha},\lambda, \eta) \equiv
  \bar{\alpha}{Q \over k!} \int_{0}^{1}\!{\dd r\over r}\, \left({\eta r/\mu^{*} \over
    1-(1-\eta)r/\mu^{*}}\right)^{k} \left({1-r/\mu^{*}\over 1-(1-\eta)
    r/\mu^{*}}\right)^{\!{\bar{\alpha}\over r}}\,\prod_{j=1}^{k-1}\left({\bar{\alpha}\over r}+j\right).
  \label{CKS_II}
\end{equation}
The total sampled clone count can also be explicitly expressed the
integral over $C^{\rm s}(\bar{\alpha},r,\lambda\vert \eta)$ from
Eq.~\ref{CS}:

\begin{equation}
  C^{\rm s}(\bar{\alpha},\lambda, \eta) = Q
  \int_{0}^{1}\!\dd r\,\left[1-\left({1- r/\mu^{*}\over 1-(1-\eta)
      r/\mu^{*}}\right)^{\!{\bar{\alpha}\over r}}\right].
  \label{CT_II}
\end{equation}


\subsubsection*{Clone-specific immigration, fixed $r$:
  $\pi(\alpha,r) = \pi_{\alpha}(\alpha)\delta(r-\bar{r})$}

Finally, using the same rate dimensionalization as before
(Eqs.~\ref{NEUTRAL_DIMENSION}), we find explicitly

\begin{equation}
  c_{k}^{\rm s}(\bar{\alpha},\lambda, \eta) =
 {2Q\over k!} \left({\eta\lambda \over
    \eta\lambda + 2\bar{\alpha}}\right)^{\!k} \sum_{j=1}^{J}{b_{j}\over C_{\star}}
  \alpha_{j}\left({2\bar{\alpha} \over
    \eta\lambda+2\bar{\alpha}}\right)^{\!\!2\alpha_{j}}\,\, \prod_{\ell=1}^{k-1}(2\alpha_{j}+\ell),
  \label{CKS_III}
\end{equation}
where here, the $\bar{\alpha}$ dependence also implicitly arises in
$\alpha_{j}$ through Eq.~\ref{ALPHAJ}. As in Eq.~\ref{CKS_I}, the
factor of $2\alpha_{j}$ arises from our choice of
nondimensionalization using $2\bar{r}$, twice the naive T cell
proliferation rate. Similarly, the total sampled clone count can be
explicitly expressed as

\begin{equation}
  C^{\rm s}(\bar{\alpha}, \lambda, \eta) =  Q\sum_{j=1}^{J}{b_{j}\over C_{\star}}
  \left[1-\left({2\bar{\alpha}\over \eta\lambda +
      2\bar{\alpha}}\right)^{2\alpha_{j}}\right].
  \label{CT_III}
\end{equation}

Note that here, $\bar{\alpha}$ is interpreted as the 
mean immigration rate over the $C_{\star}$ different clones.


\section{Small $\bar{\alpha}$ approximation}
\label{app:small_alpha}

We show that if the support of $\alpha$ is sufficiently small, terms of order
$\alpha/r$ can be neglected in the equation for $c_{k}^{\rm s}$. While
$\alpha$ is summed or integrated over, for reasonable
distributions $\pi_{\alpha}(\alpha)$, the lowest few rates contribute
the most and the average over $\pi_{\alpha}(\alpha)$ can be replaced
by a single effective value $\bar{\alpha}$. Even though $r$ is
integrated over $(0,1)$, and the region near $0^{+}$ would make
$\alpha/r$ large, the contribution from the $c_{k}^{\rm s}(\alpha, r,
\lambda, \eta)$ is also small near $r=0$. We have numerically checked
that for all cases of $\bar{\alpha} \ll 1/2$, we can approximate
$c_{k}^{\rm s}$ by

\begin{equation}
  c_{k}^{\rm s}(\alpha, r, \lambda, \eta)
  \approx {\alpha Q \over r k}\left({\eta r/\mu^{*}\over
    1-(1-\eta)r/\mu^{*}}\right)^{k}.
\end{equation}
This form is derived by approximating the product term as $\sim
(k-1)!$ and the exponential term $(\,\cdot\, )^{\alpha/r} \sim 1$.
Thus, $f_{k}^{\rm s}$ can also be approximated by

\begin{equation}
  f_{k}^{\rm s}(\bar{\alpha}, \lambda, w, \eta) = {\bar{\alpha} \over \eta \lambda w}
  \int_{{1\over 2}-{w\over 2}}^{{1\over 2}+{w\over 2}} \!\left({\eta r/\mu^{*}\over
    1-(1-\eta)r/\mu^{*}}\right)^{k} {\dd r\over r},
  \label{FKSAPPROX}
\end{equation}
where $\lambda \equiv N^{*}/Q$ and $\mu^{*}$ is given by

\begin{equation}
  \mu^{*} = {\left({1\over 2}+{w\over 2}\right)
    e^{\lambda w/\bar{\alpha}}-\left({1\over 2}-{w\over 2}\right)\over e^{\lambda w/\bar{\alpha}}-1}.
\label{MUSTAR2}
\end{equation}
Note that the $\bar{\alpha}$ prefactor in Eq.~\ref{FKSAPPROX} derived
from $\int_{0}^{\infty}\alpha \pi_{\alpha}(\alpha)\dd \alpha$ or
$\sum_{j} (b_{j}/C_{\star}) \alpha_{j}$. Since the rest of the
integrand is independent of $\alpha$, the irrelevance of the shape of
$\pi_{\alpha}(\alpha)$ is apparent.  Only the mean value
$\bar{\alpha}$ arises in this approximate calculation of $f_{k}^{\rm
  s}$.

We have explicitly shown that for small $\bar{\alpha} \ll 1$, this
approximation is quantitatively accurate. This simpler form speeds up
our numerical analysis and fitting to data. The parameters to infer
thus $\theta = \{\bar{\alpha}, \lambda = N^{*}/Q, w\}$. Even though
$\eta$ is uncertain, it is in principle controlled by the experiment.

Although we evaluate the error between the model and date as a
function of the parameters $\theta$, we have not implemented priors
for these parameters. We simply investigated the structure of the
error $H(\theta)$ and considered physiologically realistic regions in
which $\bar{\alpha}, \lambda \ll 1$.


\section{Correlated immigration $\alpha$ and proliferation $r$.}
\label{app:correlated}

\begin{figure}
\includegraphics[width=6.9in]{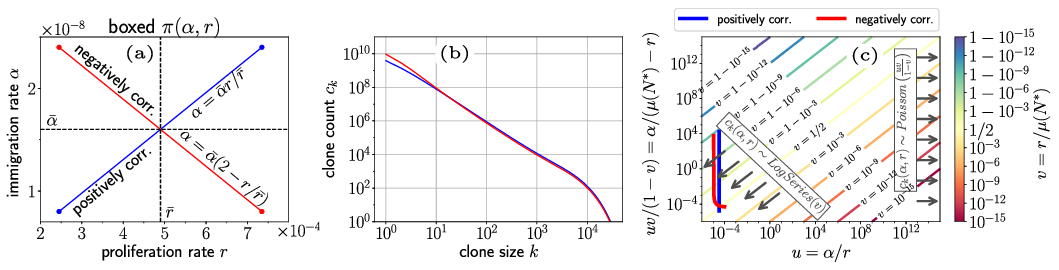}
\vspace{2mm}
\caption{Positively and negatively correlated $\pi(\alpha,r)$. (a)~For
  $\bar{r}/2\leq r\leq 2\bar{r}$, we consider $\pi(\alpha,r)$
  distributions with positively and negatively correlated $\alpha$ and
  $r$ (Eqs.~\ref{PN_PI}).  (b)~Mean sampled clone counts corresponding
  to positively and negatively correlated $\pi(\alpha,r)$ show
  negligible differences. (c)~``Line integrals'' of the positively and
  negatively correlated distributions $\pi(\alpha,r)$ in the
  $uv/(1-v)$-$u$ diagram.  Clones counts predicted by such
  $\pi(\alpha,r)$ follow log-series distributions, similar to those of
  a neutral model.}
\label{fig:correlations}
\end{figure}

Hitherto, we have considered independent immigration and
proliferation, and assumed a factorisable rate distribution
$\pi(\alpha,r) = \pi_{\alpha}(\alpha)\pi_{r}(r)$.  However,
immigration and proliferation rates may be correlated for certain
clones. For example, a frequent realization of V(D)J recombination may
also result in a TCR that is more likely to be activated for
proliferation.  In this case, $\alpha$ would be positively correlated
with $r$. In Fig.~\ref{fig:correlations} we consider the effect of
correlated $\pi(\alpha,r)$.  For $\bar{r}/2\leq r \leq 2\bar{r}$, we
considered normalized, positively/negatively correlated box
distributions as shown in Fig~\ref{fig:correlations}(a):

\begin{align}
\text{Positively correlated}:\quad  &  \pi(\alpha,r)
={1\over \bar{r}}\delta\left(\alpha-{\bar{\alpha}\over \bar{r}}r\right), \nonumber \\
\text{Negatively correlated}:\quad  & \pi(\alpha,r) = {1\over \bar{r}}
\delta\left(\alpha-\bar{\alpha}\left(2- {r\over \bar{r}}\right)\right).
\label{PN_PI}
\end{align}
%
Within our mean field model, these correlated distributions
$\pi(\alpha,r)$ result in very similar expected clone abundance
distributions $c_k$ (Fig~\ref{fig:correlations}(b)).  This
insensitivity to correlations between immigration and proliferation
can be qualitatively understood by considering the ``line integral''
over dominant paths of $\pi(\alpha,r)$ in the
$uv/(1-v)=\alpha/(\mu(N^*)-r)$ \textit{vs.} $u=\alpha/r$ diagram. As
shown in Fig.~\ref{fig:correlations}(c), both line integrals remain in
the log-series distribution regime, indicating that the clone
abundance distributions are qualitatively similar to that predicted by
a model with proliferation heterogeneity alone.

\section{$c_{k}$ for clustered immigration}
\label{app:clustered}

We explore how clustered emigration from the thymus affects the mean
clone count $c_{k}$. Suppose that $q$ cells of the same clone (TCR
nucleotide or amino acid sequence) are simultaneously exported by the
thymus.  The equation for the mean clone count $c_{k}$ is thus

\begin{equation}
{\dd c_k \over \dd t} = \sum_{q}\alpha_{q}\left[c_{k-q}
-c_k\right] + r\left[(k-1)c_{k-1}-kc_k\right]
+\mu(N)\left[(k+1)c_{k+1}-kc_k\right].
\label{CK_ODE_Q}
\end{equation}
This equation does not admit a simple analytic solution so we
numerically solved the equation assuming $\alpha_{q} =
\alpha_{5}\mathds{1}(q,5)$ and $Q=10^{11}$. The shape of $c_{k}$ for
single cell immigration ($q=1$) is compared to that of $q=5$.
\begin{figure}
\begin{center}
\includegraphics[width=3.1in]{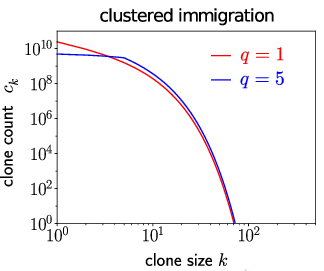}
\end{center}
\caption{\label{CK_CLUSTER} Clustered immigration in a neutral model.
  Comparison of clone abundances for a $q=1$ and $q=5$ models.  The
  difference between the two predicted mean clone counts arise for $k
  \lesssim q$. Even after sampling, clone counts predicted under
  clustered immigration ($q>1$) yields a more slowly decreasing
  $c_{k}^{\rm s}$ for small $k \lesssim q$.}
\end{figure}
For $q>1$, $c_{k}$, and ultimately $c_{k}^{\rm s}$ and $f_{k}^{\rm s}$
is flatter up to $k \approx q$, making the clone counts kink more
downwards near $q$. Thus, as can be seen from Fig.~\ref{FK_FIT}(a,b),
we can reasonably conclude that some level of paired immigration would
provide even better fits to the data at appropriately small values of
$\lambda$, especially for the first few $k$-points.

\end{document}